%% file: dissertation.tex
\pdfoutput=1
\documentclass[11pt,a4paper]{report}

\usepackage{mmp}
\usepackage{booktabs}

\usepackage{cite}

\usepackage{caption}

\begin{document}


\pagenumbering{roman}

\include{cover}

\pagestyle{empty}


\include{abstract}                 

\pagenumbering{roman}
\pagestyle{fancy}
\fancyhead{}
\fancyfoot[C]{\thepage}
\renewcommand{\headrulewidth}{0 pt}
\renewcommand{\chaptermark}[1]{\markboth{#1}{}}

\tableofcontents   
\newpage
\listoffigures
\newpage 
\listoftables
\newpage

\pagenumbering{arabic}

\setchapterheaderfooter

\include{Chapter1/chapter1}
\include{Chapter2/chapter2}
\include{Chapter3/chapter3}
\include{Chapter4/chapter4}
\include{Chapter5/chapter5}
\include{Chapter6/chapter6}
\include{Chapter7/chapter7}

\newpage  
\thispagestyle{plain}  
\input{acknowledgements}

\pagestyle{bibliography}

\begingroup
\renewcommand{\addcontentsline}[3]{}
\input{References/ref.tex}  
\endgroup

\end{document}

%% file: cover.tex
\title{A Benign Activity Extraction Method for Malignant Activity Identification using Data Provenance}

\author{Taishin Saito}

\authoremail{mgs231002@iisec.ac.jp}

\degreeschemecode{} 
\degreeschemetitle{Informatics} 
\degreetype{MSc}

\modulecode{} 
\moduletitle{} 

\date{28th February 2025} 

\status{Final} 
\version{1.0}

\supervisor{Dr./Prof. Kuniyasu Suzaki} 

\supervisoremail{suzaki@iisec.ac.jp}

\maketitle

%% file: abstract.tex
\thispagestyle{empty}

\begin{center}
    {\LARGE\bf Abstract}
\end{center}

In order to understand the overall picture of cyber attacks and to identify the source of cyber attacks, a method to identify malicious activities by automatically creating a graph that ties together the dependencies of a series of related events by tracking Data Provenance has been developed. However, the problem of dependency explosion, in which a large number of normal computer system operations such as operations by authorized users are included in the dependencies, results in a huge generated graph, making it difficult to identify malicious activities.

In this paper, we propose a method to reduce the search space for malicious activities by extracting and removing frequently occurring benign activities through natural language processing of log data and analysis of activities in the computer system using similarity judgments. In the evaluation experiment, we used the DARPA TC Dateset, a large-scale public dataset, to evaluate the effectiveness of the proposed method on the dependency explosion problem. In addition, we showed that about 6.8 to 39\% of the activities in a computer system could be defined as patterns of benign activities. In addition, we showed that removing benign activities extracted from a portion of the log data (approximately 1.4\% to 3.2\% in size) can significantly reduce the search space (up to approximately 52\%) in large data sets.

%% file: Chapter1/chapter1.tex
\chapter{Introduction}

\section{Background}
\begin{figure}[tb]
\begin{center}
\includegraphics[scale=0.8]{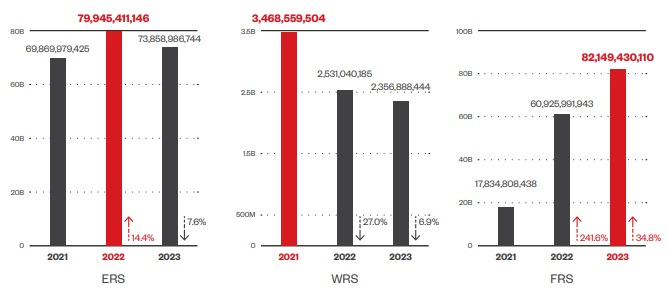}
\end{center}
\caption{Ransomware Detection Layers (Left: Email, Center: Web, Right: File)\cite{TrendMicro}}
\label{fig:threat-layer}
\end{figure}

In recent years, cyber attacks on organizations have been becoming increasingly complex and sophisticated. In particular, the damage caused by targeted attacks and ransomware has been escalating, making it difficult to counter such threats using traditional defense mechanisms alone. Under these circumstances, as cyber attacks continue to advance, efforts to completely prevent intrusions by attackers are facing significant challenges. Figure \ref{fig:threat-layer} illustrates the number of ransomware detections by detection layer, based on data published in 2023 \cite{TrendMicro}. Traditionally, ransomware was often detected through emails or the web, suggesting that ransomware attacks were predominantly conducted via large-scale spam email campaigns containing malicious links or other bulk-distribution methods over email and the web. However, in recent years, it has been reported that a significant portion of ransomware detections occur as files within endpoints, indicating that attackers have successfully breached initial defenses, allowing ransomware to reach deeper into computer systems \cite{TrendMicro}.

\begin{figure}[tb]
\begin{center}
\includegraphics[scale=0.6]{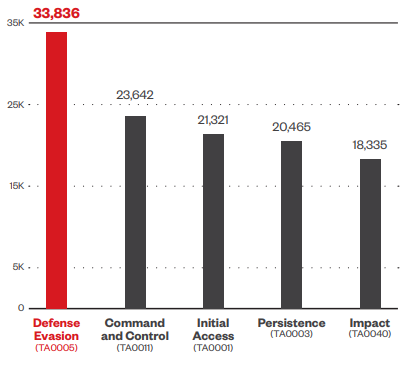}
\end{center}
\caption{Detection Counts by MITRE ATT\&CK Tactics\cite{TrendMicro}}
\label{fig:tactics}
\end{figure}

One of the factors contributing to these intrusions is the use of customized attack techniques in advanced targeted attacks, which enable attackers to evade intrusion detection systems \cite{APT}. Figure \ref{fig:tactics} shows the number of attacks detected in 2023, categorized by MITRE ATT\&CK tactics \cite{TrendMicro}. Among the tactics defined in MITRE ATT\&CK\cite{MITRE}, which is a framework for classifying attackers’ strategies, techniques, and procedures, those classified under "Defense Evasion" were detected the most frequently. This suggests that attackers are focusing their efforts on evading detection by security tools. Furthermore, in recent years, there has been an increase in interactive attacks conducted manually by human operators without relying on malware, leading to an increase in stealthy and sophisticated attacks that are difficult to detect \cite{threat_report}.

Another major factor that allows intrusions is the inability to effectively analyze the vast number of alerts generated daily by intrusion detection systems. Since it is difficult to completely prevent intrusions, methods for promptly investigating the root cause of an attack once an intrusion has been identified are actively being explored.

When an attack is detected, investigations are conducted using log collection and forensic techniques. However, there is a shortage of security professionals with the necessary technical expertise to analyze the overwhelming number of alerts generated by intrusion detection systems, often exceeding the number that analysts can handle \cite{SOC}\cite{workforse_report}.

Given the limitations of manual analysis by security analysts, Data Provenance-based analysis is increasingly being adopted to facilitate rapid investigation of attack activities \cite{p-EDR-interview}. Data Provenance analysis automatically correlates event dependencies within a computer system based on log data and outputs them as a graph. This allows security teams to visualize a sequence of events as a continuous attack activity, thereby simplifying the previously manual analysis process \cite{Muhammad}. In other words, by visualizing the dependencies of processes, files, and other entities as a dependency graph, analysts can conduct investigations more efficiently.

Furthermore, to detect attacks that traditional detection systems—based on individual files, processes, and other isolated entity information—fail to recognize, provenance-based Endpoint Detection and Response (EDR) systems have been developed, utilizing dependency relationships between entities within a computer system to enhance attack detection \cite{PROGRAPHER}\cite{SLEUTH}\cite{UNICORN}\cite{ATLAS}\cite{nodlink}.

However, a major challenge in utilizing Data Provenance is the "dependency explosion" problem. When linking events within a computer system using simple rule-based correlation, benign activities unrelated to the attack may also be included in large volumes, leading to the generation of enormous graphs. Ideally, the generated graph should consist solely of attack-related activities, but due to the dependency explosion issue, analysts must analyze massive graphs that contain both benign and malicious events, imposing a significant burden on their workload.

Therefore, it is necessary to address the dependency explosion issue in Data Provenance-based graphs and generate optimized graphs where unnecessary dependencies have been removed while preserving attack-related dependencies.

Numerous approaches have been proposed to mitigate dependency explosion. Existing methods primarily focus on identifying and excluding individual benign events from the analysis or extracting malicious activities \cite{PROGRAPHER}\cite{LogGC}\cite{CPR}\cite{DEPIMPACT}\cite{NODOZE}. However, identifying individual benign events is challenging because the classification of an event as benign depends on an organization’s policies and the specific characteristics of its computer systems, making it difficult to apply universally across all systems. Additionally, extracting only malicious activities requires continuously updating knowledge to keep up with evolving attack techniques. Even when anomaly-based detection methods that do not rely on predefined malicious activity patterns are used, they still require periodic retraining to accommodate changes in benign behavior, such as login attempts to newly deployed legitimate servers.

\clearpage

\section{Purpose and Contribution}
The objective of this study is to address the issue of dependency explosion in order to facilitate the analysis of attack activities. Furthermore, this study aims to develop a method that can adapt to the unique activity patterns of an organization's computer system while reducing dependencies without requiring frequent retraining.

To achieve this, we propose a method that leverages natural language processing and similarity analysis to extract benign activities within a computer system and remove them from the dependency graph.

Contributions of This Study
The contributions of this study are as follows:

We propose a method to extract frequently occurring benign activities from log data and remove them from the dependency graph. To the best of our knowledge, this is the first attempt worldwide to address the issue of dependency explosion in this manner.
We demonstrate that by removing benign activities extracted from approximately 3\% of the log data, dependency relationships can be reduced by up to 52.3\%. This result highlights the effectiveness of the proposed method.
We show that approximately 6.8\% to 39\% of activities within a computer system can be defined as patterned benign activities. This insight supports the practical feasibility of benign activity extraction.
We demonstrate that frequently occurring benign activities can be extracted from a small portion (approximately 11\% or less) of the analyzed log data. By removing these benign activities from the dependency graph, the search space within large-scale log data can be significantly reduced.
We analyze the publicly available DARPA Transparent Computing (TC) Data and estimate its characteristics. Additionally, to ensure accuracy in identifying malicious nodes within the dataset, we consulted Shaofei Li (Peking University), one of the authors of a previous study that utilized this data \cite{nodlink}.
This research has been presented at the Computer Security Symposium 2024 (CSS 2024), organized by the Information Processing Society of Japan (IPSJ) \cite{CSS}.

\section{Structure of This Paper}

This paper is structured as follows:

\begin{itemize}
\item \textbf{Chapter 1: Introduction}
This chapter clarifies the background of increasingly sophisticated cyber attacks, identifies challenges and issues in existing attack analysis and detection methods, and examines the current trends and challenges in the utilization of Data Provenance. Based on this, it defines the objectives and contributions of this study. Additionally, the structure of this paper is outlined.\\

\item \textbf{Chapter 2: Utilization of Data Provenance} \\ This chapter defines Data Provenance and discusses its application in malicious activity analysis. Furthermore, it organizes the current challenges, focusing on the dependency explosion problem and previous research. Based on this, the research direction of this study is presented. \\ 

\item \textbf{Chapter 3: Proposed Method} \\ Taking into account the threat model and assumptions underlying this research, this chapter details the proposed method. The explanation is divided into four phases, from reading log data to outputting a dependency-reduced graph.\\

\item \textbf{Chapter 4: Evaluation Experiments} \\ This chapter presents the conditions of the evaluation experiments conducted in this study. The experimental conditions include the experimental environment, parameter settings, evaluation metrics, and dataset details. It also describes the characteristics and analysis results of the \textbf{DARPA Transparent Computing (TC) Data}, which was used as the dataset. The results of the conducted experiments are then presented. \\ 

\item \textbf{Chapter 5: Discussion} \\ Based on the experimental results, this chapter discusses the effectiveness of dependency reduction achieved through benign activity extraction. Furthermore, it describes the characteristics of the analyzed data and the proportion of extracted benign activities. Additionally, it outlines the limitations of the proposed method and discusses the current challenges. \\ 

\item \textbf{Chapter 6: Related Work} \\ This chapter reviews related studies that address the dependency explosion problem. It categorizes existing approaches into two types: methods that exclude individual benign events from logs and methods that focus on identifying malicious activities. Based on this, the differences between these approaches and the present study are highlighted. \\ 

\item \textbf{Chapter 7: Conclusion} \\ This chapter summarizes the effectiveness of the proposed method in relation to the background and objectives of this research, as well as its correlation with data characteristics. Additionally, it organizes unresolved issues and discusses future research directions. \\

\end{itemize}

%% file: Chapter2/chapter2.tex
\chapter{Utilization of Data Provenance}

In this chapter, we describe the definition of Data Provenance, its linkage to malicious activities, the issue of dependency explosion, prior research, and the positioning of this study.

\section{Data Provenance}

Data Provenance represents the origin of data and tracks the record of processes that led to its current state \cite{SP800}. In other words, it refers to records that trace where data originated, how it was used, and where it was transmitted. Within a computer system, Data Provenance includes information needed to track who accessed which data and when, as well as to identify the reasons why a file reached its current state. This information, which captures the origin and history of data throughout its lifecycle, is generated and stored based on audit logs, event logs, and other log data \cite{DataProvenance}. The provenance graph generated using Data Provenance is represented as a directed acyclic graph (DAG), where system entities are represented as nodes and system calls used in events between nodes are represented as edges \cite{p-EDR}. System entities include processes, files, hosts, and other relevant components.

\begin{figure}[tb]
\centering
\includegraphics[scale=0.4]{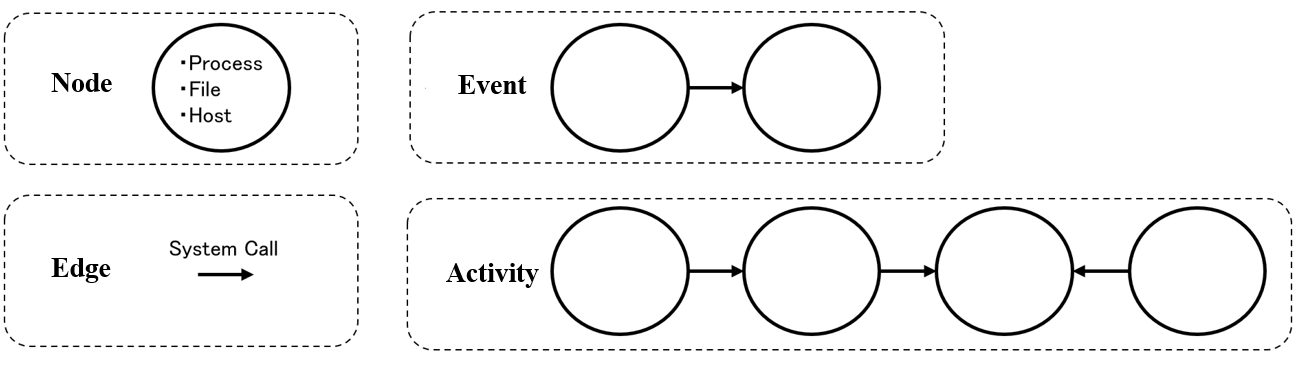}
\caption{Event \& Activity}
\label{fig:def}
\end{figure}

\begin{figure}[tb]
\centering
\includegraphics[scale=0.6]{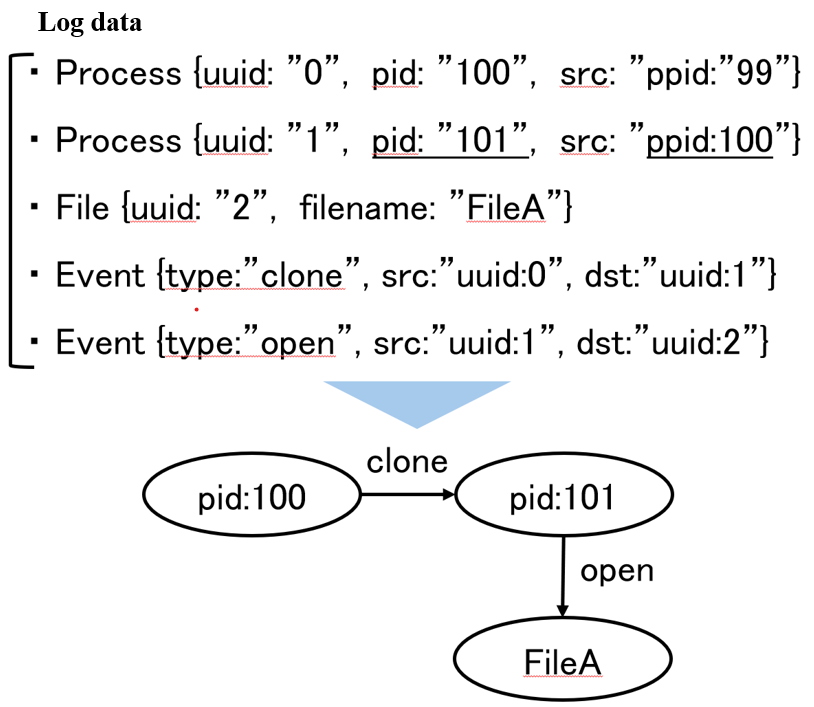}
\caption{Log data \& Provenance Graph}
\label{fig:log-graph}
\end{figure}

Furthermore, in this study, as shown in Figure \ref{fig:def}, we define a unit consisting of one edge and two nodes in the provenance graph as an "event," while a unit composed of a sequence of multiple edges is referred to as an "activity." The example in Figure \ref{fig:log-graph} illustrates an activity comprising three nodes, where two system calls, \texttt{clone} and \texttt{open}, are represented as edges.

\section{Linking Malicious Activities}

When detecting an event suspected to be part of a cyber attack, it is difficult to determine whether it is related to malicious activity by focusing on a single event alone. For example, an event such as sending an email can be used for attacks, but it is also commonly observed in normal communication, making it difficult to determine whether the event is attack-related. Whether an event is linked to an attack should be assessed in the context of a series of activities, requiring individual events to be correlated and represented as part of an activity.

As a method for correlating attack-related logs, \textbf{BackTracker} was proposed, which visualizes the dependency relationships of processes and files that may have influenced the detection point using a graph \cite{Backtracker}. BackTracker obtains dependencies between processes, between processes and files, and between processes and filenames. Detection points typically use events detected by existing intrusion detection systems. By performing causal analysis backward from the detection point, it is possible to determine the root cause of the attack, while forward causal analysis can identify the scope of its impact. By automating causal analysis, dependency graphs can be constructed to explain relationships between events without manually correlating individual log entries from the analyzed log data.

Audit logs and event logs are used to correlate processes and files within a computer system, linking them based on process IDs, file paths, and event information recorded in the collected log files.  
Figure \ref{fig:log-graph} illustrates the process of converting audit log data into a provenance graph. The upper part of the figure represents raw log data, containing the following three key pieces of information:

\begin{itemize}
\item \textbf{Process Information:} Two processes (PID:100, PID:101) and their parent-child relationship.
\item \textbf{File Information:} Attributes of the target file (FileA).
\item \textbf{Event Information:} Records of interactions between processes and files (e.g., open operation).
\end{itemize}

These pieces of information are converted into a provenance graph, as shown in the lower part of the figure, in the following manner:

\begin{itemize}
\item \textbf{Nodes:} Processes and files are represented as individual nodes.
\item \textbf{Edges:} Parent-child relationships between processes and system calls related to file operations are represented as edges.
\item \textbf{Temporal Order:} The chronological order of events is preserved in the graph structure.
\end{itemize}

This transformation makes it easier to visually understand how individual log entries are related. The generation and utilization of provenance graphs allow tracking of a sequence of activities occurring within a computer system. Such graphs are particularly useful when investigating the origin of an event or assessing the impact of an event on the system.

\begin{figure}[tb]
\centering
\includegraphics[scale=0.4]{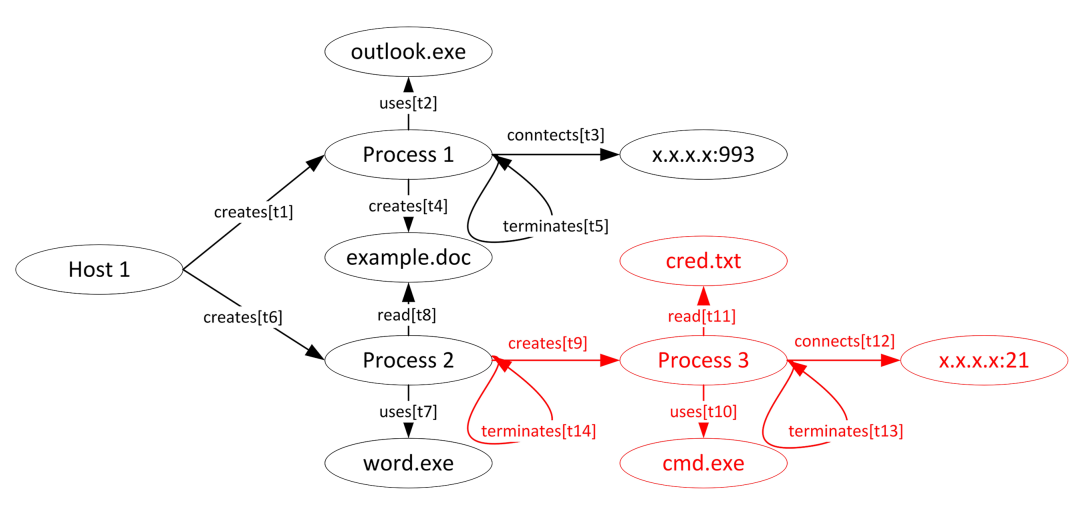}
\caption{Tracking Malicious Activities Using Provenance Graphs\cite{p-EDR}}
\label{fig:Provenance-graph}
\end{figure}

An example of tracking malicious activities using a provenance graph is illustrated in Figure \ref{fig:Provenance-graph} \cite{p-EDR}. In this figure, the activity of a phishing email attack is depicted, where a user views an email and executes an attached file. First, Process 1 connects to a mail server (x.x.x.x:993) using Outlook and retrieves an attachment named \textit{example.doc}, which contains malicious code. The execution of \textit{example.doc} is performed by the victim user. Next, the execution of \textit{example.doc} generates Process 2, which handles document processing and uses Microsoft Word. At this stage, the malicious code is executed as a Word macro within Process 2, which then spawns a malware process, Process 3. The malware activity, indicated in red, shows that Process 3 reads sensitive information from \textit{cred.txt} using the command prompt and transmits it externally (x.x.x.x:21).

\section{Challenge: Dependency Explosion}

\begin{figure*}[htpb]
\centering
\includegraphics[width=\hsize]{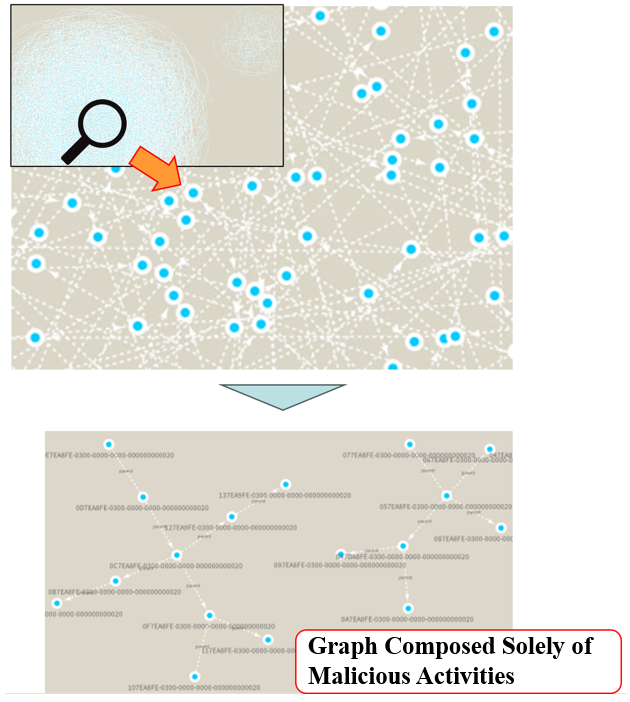}
\caption{Dependency Explosion and Dependency Reduction Visualization}
\label{fig:dependency-explosion}
\end{figure*}

The linkage of malicious activities using Data Provenance faces a significant challenge known as dependency explosion, where an excessive number of dependencies are generated in the provenance graph \cite{suetugu}\cite{miyasaka}. The phenomenon of dependency explosion is illustrated in Figure \ref{fig:dependency-explosion}. In dependency analysis using Data Provenance, graphs with over a million edges can be generated, requiring analysts to extract malicious activities from massive graphs. This contradicts the goal of reducing analysts’ workload.

The main factors contributing to dependency explosion include the massive size of the audit log data being analyzed and the use of simple correlation rules, such as parent-child relationships between processes, which can introduce dependencies unrelated to actual attacks. The audit logs generated by modern, large-scale, and complex systems continue to grow daily, and depending on the analysis period, it may be necessary to process tens of terabytes of data. Additionally, dependencies generated by simple rules often include routine system operations and normal user activities, making it difficult to filter out attack-related events effectively.

To efficiently identify the origin and impact of malicious activities, it is essential to extract only the dependencies that are highly relevant to attacks and generate a graph composed solely of attack-related nodes and edges. A survey of technical managers handling EDR (Endpoint Detection and Response) products indicated that, considering workload constraints, the number of edges output in a provenance graph should ideally be limited to around 10 to 100 \cite{p-EDR-interview}.

\section{Existing Approaches and Challenges}

Various methods have been proposed to address the issue of dependency explosion. Existing approaches can mainly be classified into the following two categories.

\subsection{Exclusion of Benign Events}

Some methods focus on identifying individual events that are generally considered benign and excluding them from the analysis \cite{LogGC}\cite{CPR}. By preemptively removing known benign events from the log data being analyzed, the size of the generated graph can be reduced, thereby mitigating dependency explosion.

In the LogGC method proposed by Lee et al., the focus was placed on events related to the deletion of temporary files, which were found to be excessively present in log data \cite{LogGC}. Since the deletion of temporary files does not affect subsequent system operations and is not useful for attack analysis, these events were excluded from the analysis. This approach reduced the analyzed log data by approximately 23.8\%, thereby alleviating dependency explosion.

Additionally, the CPR method proposed by Xu et al. reduces logs by merging duplicate events \cite{CPR}.

\subsection{Identification of Malicious Activities Using Graph Weighting}

Another approach involves assigning weights to edges and nodes to extract activities with a high likelihood of being malicious, keeping only the malicious activities in the graph \cite{NODOZE}\cite{DEPIMPACT}\cite{nodlink}. By retaining only the extracted malicious activities, smaller graphs are generated, thereby mitigating dependency explosion.

In the DEPIMPACT method proposed by Fang et al., data flow volume, timing, and node degree were used as weighting indicators \cite{DEPIMPACT}. Events with a data flow volume or timestamp similar to the detection point, as well as events that have a high ratio of outgoing accesses relative to incoming accesses from other nodes, were identified as highly attack-related events. By generating a graph consisting solely of identified malicious activities, dependency explosion was reduced.

\subsection{Summary and Challenges of Existing Approaches}

To address dependency explosion, methods have been proposed that either define rules to exclude individual benign events or nodes or focus on identifying sequences of malicious activities.

However, excluding a single benign node raises concerns about whether it could potentially be exploited in an attack, making it uncertain whether it should be completely removed as unrelated to attacks. Furthermore, even if an event is generally considered benign, differences in organizational policies and computer system configurations may mean that certain information remains necessary for analysis.

Additionally, identifying malicious activities requires continuously updating attack patterns whenever new attacks emerge. Moreover, to accommodate changes in benign activities due to user behavior or system updates, periodic retraining may be required.

\section{Proposed Approach}

To effectively address dependency explosion, a solution is needed that does not require frequent retraining due to new attacks or changes in benign activities and is not dependent on specific environments.

Thus, we hypothesized that by acquiring patterns of benign activities, it would be possible to identify log data that can be consistently excluded from analysis. If an organization retains patterns of benign activities unique to its computer system, it may be possible to reduce the amount of log data analyzed at the activity level. Since the extracted patterns are derived from the target system itself, the method can be applied universally across different systems. Furthermore, benign activities are presumed to be more stable than malicious activities, suggesting that frequent retraining may not be necessary.

Based on these considerations, this study proposes a method that leverages benign activity patterns extracted from the analyzed data to address dependency explosion without requiring frequent retraining.

%% file: Chapter3/chapter3.tex
\chapter{Proposed Method}

In this chapter, we describe the proposed method for extracting benign activities within a computer system and removing the extracted activities from the dependency graph. First, we present the threat model assumed in this study, followed by a detailed explanation of the proposed method.

\begin{figure*}[tb]
\centering
\includegraphics[scale=0.4]{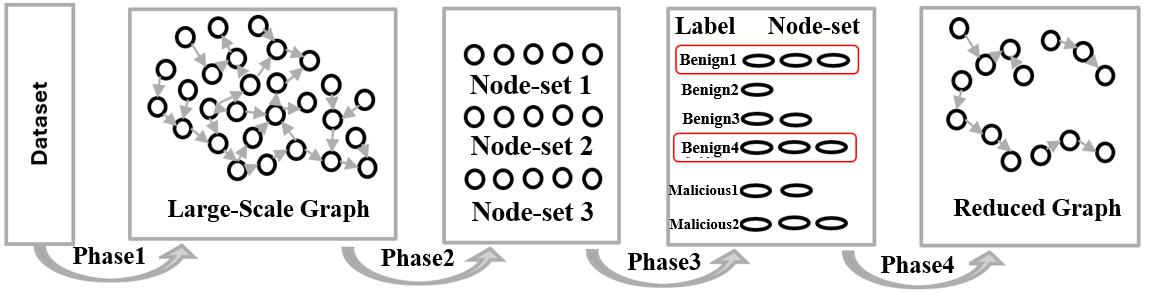}
\caption{Overview of the Proposed Method}
\label{fig:overview}
\end{figure*}

An overview of the proposed method is illustrated in Figure \ref{fig:overview}. The proposed method consists of four phases: data preprocessing, node-set construction, node-set labeling, and benign activity extraction along with dependency graph reduction.

\section{Threat Model and Assumptions}

The threat model assumed in this study follows prior research, where it is assumed that log collection by the audit system operates correctly and records sufficiently detailed system activities \cite{DEPIMPACT}\cite{NODOZE}\cite{nodlink}. Additionally, we assume that the storage and transmission of log data are securely managed, and we do not consider scenarios where logs have been tampered with. Furthermore, attacks that are not reflected in the logs, such as side-channel attacks, are outside the scope of this study.

Moreover, as discussed in Section \ref{addlabel}, when using labeled data for extracting benign activities, it is necessary to accurately identify which nodes are malicious in cases where the data contains malicious nodes. However, datasets that do not contain any malicious nodes are also acceptable as labeled data.

\section{Phase 1: Data Preprocessing}
\label{preprocess}

When constructing a dependency graph of events within a computer system, it is necessary to extract system calls used in events from log data and obtain the corresponding nodes related to these edges. In our proposed method, following prior research, we acquire processes, file objects, and net flow objects as nodes \cite{DEPIMPACT}\cite{NODOZE}\cite{nodlink}.

For the mechanism that retrieves relevant information from log data, we integrate THREATRACE, a host-based anomaly detection system, for threat analysis \cite{Threatrace}\cite{Threatrace-git}.

Additionally, the analyzed events were classified based on system calls. The system calls included in the analysis are listed in Table \ref{tab:syscall}. Events occurring between processes and files, between processes, and between processes and network flows were collected based on their corresponding system calls. The selected system calls were those used in previous studies and were present in the dataset used in this research \cite{DEPIMPACT}\cite{nodlink}. Although events involving system calls not listed in Table \ref{tab:syscall} may exist, they were excluded from the analysis and removed during data preprocessing.

\begin{table}[t]
\centering
\caption{Analysis Target}
\label{tab:syscall}

  \begin{tabular}{cc}
   \hline
    Events&System Call\\
   \hline  
   \hline 
   Process/File&open, read, write, chmod, pipe\\
   Process/Process&execve, clone\\
   Process/NetFlow&recvfrom, sendto, recvmsg, sendmsg\\
   \hline
  \end{tabular}

\end{table}

\section{Phase 2: node-set Construction}
\label{createnodeset}

In this method, natural language within the log data is utilized for computation. For the natural language processing component, we integrate the NODLINK method from previous research, which converts command lines, file paths, IP addresses, and port numbers into numerical vectors using FastText \cite{nodlink}\cite{nodlink-git}\cite{fasttext}.

\begin{figure}[tb]
\centering
\includegraphics[width=\hsize]{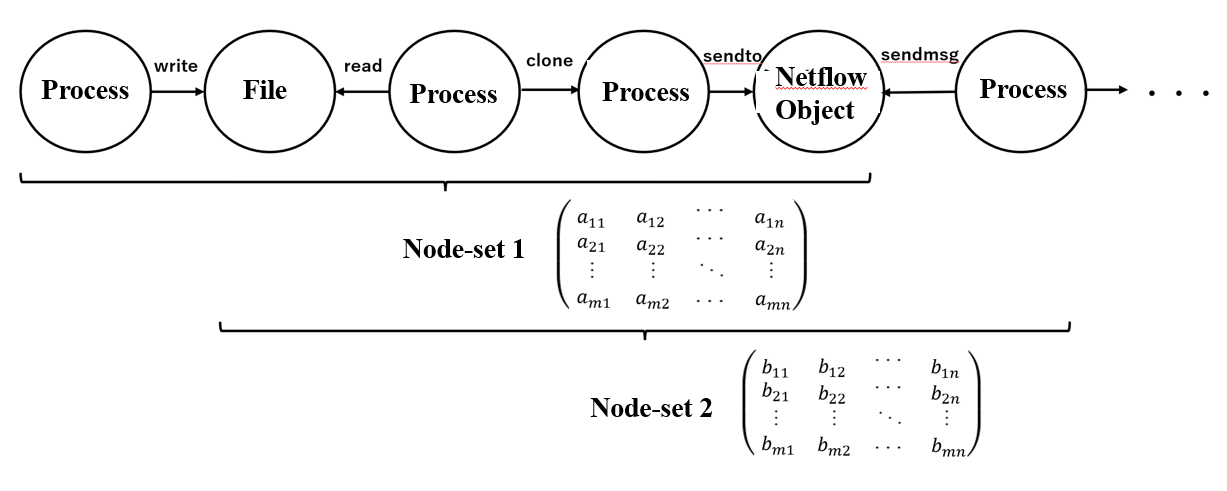}
\caption{Extracting Node-sets}
\label{fig:nodeset}
\end{figure}

Additionally, our proposed method defines node-sets by extracting portions of activities from the dependency graph. An illustration of node-set construction is shown in Figure \ref{fig:nodeset}. A node-set consists of a partial segment of an activity that is connected through event chains. The size of each node-set is defined as an activity containing five nodes.

Furthermore, to define the feature representation of a node-set, we represent the characteristics of the nodes within the node-set as a single numerical vector. The node-set features are composed of the features of the five nodes included in the node-set, and they are defined based on the approach used in NODLINK, as expressed by the following equation \cite{nodlink}.

\begin{equation}
\label{embed}
V=w_{c}*V_{c}+\sum w_{f_{i}}*V_{f_{i}}+\sum w_{n_{i}}*V_{n_{i}}
\end{equation}

Here, $ V_{c}$, $ V_{f_{i}}$, and $ V_{n_{i}}$ represent the distributed representations of command lines, files, and network flows, respectively, while $ w_{c}$, $ w_{f_{i}}$, and $ w_{n_{i}}$ are their corresponding weights. The weight of a file, $ w_{f_{i}}$, is given by:

\begin{equation}
w_{f_{i}} = \log \frac{P}{P_{f_{i}}}
\end{equation}

where $ P$ represents the total number of events, and $ P_{f_{i}}$ denotes the number of events involving a specific file $ f_{i}$. The weight for network flows follows the same calculation as files, while the weight for processes is defined as the average of the weights of files and network flows.

The weights in Equation \ref{embed} are assigned to account for files and IP addresses that are commonly used across different processes. For example, if all processes load the libc file, libc is not useful for extracting unique characteristics of individual processes. By assigning weights in this manner, the influence of files and IP addresses accessed by multiple types of processes is mitigated \cite{nodlink}.

\section{Phase 3: node-set Labeling}
\label{addlabel}

\begin{figure}[tb]
\centering
\includegraphics[width=\hsize]{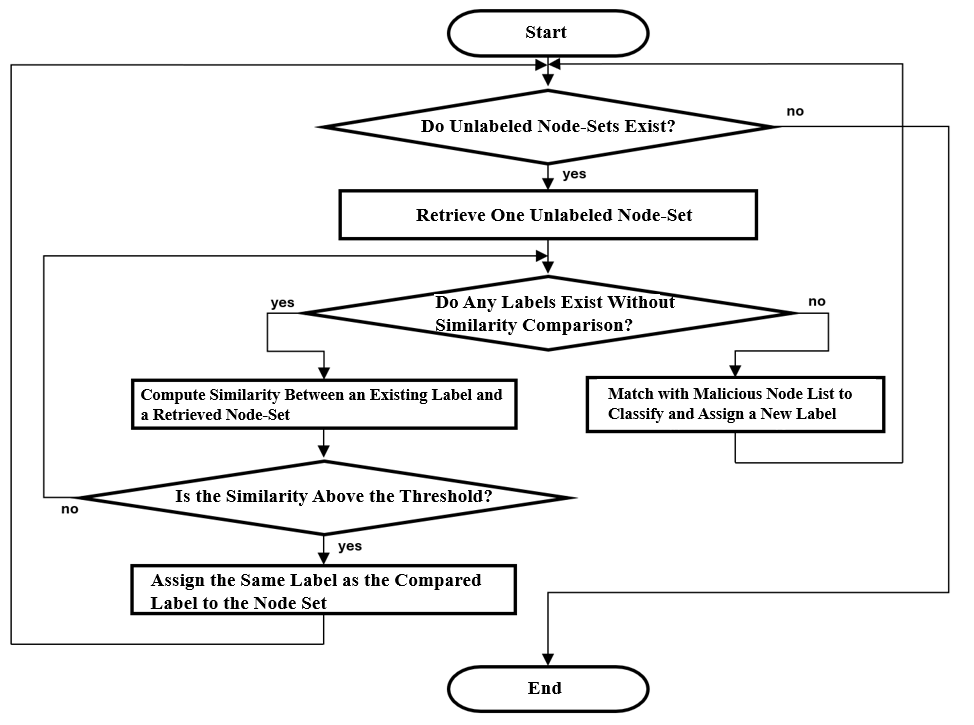}
\caption{Labeling Flow for Node-Sets}
\label{fig:flow}
\end{figure}

In this method, node-sets are labeled based on the assumption that the classification of individual nodes as either malicious or benign is known in advance. An overview of the labeling process is illustrated in Figure \ref{fig:flow}. A node-set is classified as malicious if at least one of its nodes is malicious, and as benign if all nodes within the set are benign.

Additionally, to enhance the labeling process, a mechanism is introduced that classifies node-sets with similar features under the same label by utilizing feature similarity. The labeling process begins by selecting the first node-set and determining its label by cross-referencing the nodes within the set against a list of known malicious nodes. If any of the nodes are found in the malicious node list, the set is labeled as malicious; otherwise, it is labeled as benign.

Next, when a new node-set is processed, its features are compared with those of already labeled node-sets. If the similarity is high, the new node-set is assigned the same label. Cosine similarity is used as the similarity metric for this comparison. If the similarity score falls below a predefined threshold, the node-set is cross-referenced against the malicious node list, and a new label is assigned accordingly.

By applying this process to all node-sets, multiple labels are generated within both the benign and malicious categories, allowing similar node-sets to be grouped under the same classification.

\section{Phase 4: Extraction of Benign Activities and Reduction of the Dependency Graph}
\label{removegraph}

Using the labeled node-sets created in Section \ref{addlabel}, benign activities are removed from the dependency graph.

Benign activities are identified based on the labels classified as benign. The benign labels are ranked according to the number of node-sets associated with each label, and the characteristics of the top-ranked labels are defined as benign activities. During the labeling process described in Section \ref{addlabel}, each label is assigned to one or more node-sets. If many node-sets exhibit similar characteristics to a particular label, the number of node-sets associated with that label increases. Therefore, benign labels with a large number of associated node-sets represent frequently occurring benign activities.

Finally, the extracted benign activities are removed from the graph. The removal process follows the same methodology used in Section \ref{createnodeset}, where node-sets consisting of five nodes are extracted. Each extracted activity is compared to the characteristics of the top-ranked benign labels, and if the similarity exceeds a predefined threshold, the corresponding activity is removed from the graph.

%% file: Chapter4/chapter4.tex
\chapter{Evaluation Experiments}

In this chapter, we present the experimental conditions, details and analysis of the dataset used, and the experimental results to evaluate the effectiveness of the proposed method in reducing dependency relationships.

\section{Experimental Conditions}

\noindent\textbf{Experimental Environment}:  
The experiments were conducted on a server equipped with an Intel(R) Xeon(R) Silver 4314 (16 cores, 2.4 GHz) CPU, 256 GB of memory, and running Ubuntu 22.04 64-bit OS.

\vskip.5\baselineskip
\noindent\textbf{Parameters}:  
The proposed method includes multiple parameters that could influence the experimental results. In this experiment, only the parameter defining the number of top-ranked labels considered for graph reduction was varied. We conducted experiments using values ranging from the top 3 labels to the top 1500 labels.

Other parameters were fixed: the node set size was set to 5 nodes, and the cosine similarity threshold was set to 1.0.

\vskip.5\baselineskip
\noindent\textbf{Evaluation Metrics}:  
The evaluation metrics used in this experiment include False Negatives (FN), False Positives (FP), and graph reduction rate. FN represents the number of malicious nodes that were erroneously removed from the graph. FP represents the number of benign nodes that remained in the graph. The graph reduction rate measures the proportion of nodes removed from the graph, calculated by comparing the number of nodes in the graph generated after preprocessing in Section \ref{preprocess} with the number of nodes remaining after removing extracted benign activities. Additionally, execution time was measured for the operations in Section \ref{removegraph}, which involve ranking the labels and removing benign activities from the graph using the labeled data.

\vskip.5\baselineskip
\noindent\textbf{Dataset}:  
For this experiment, we used the DARPA Transparent Computing (TC) Data dataset. The dataset was divided into labeled data for extracting benign activities and evaluation data for generating the provenance graph subject to dependency reduction. The proportion of each dataset is described in the next section.

The processing of obtaining edge and node information in Section \ref{preprocess} was applied to both the labeled data and the evaluation data, meaning that this processing was executed twice per dataset. The evaluation data was used to generate the graph for reduction in Section \ref{removegraph}, without performing the operations in Sections \ref{createnodeset} and \ref{addlabel}. Instead, the information obtained in Section \ref{preprocess} was used as input.

\section{DARPA Transparent Computing (TC) Data}

\subsection{Dataset Details}

The TC Data dataset was created as part of DARPA’s Transparent Computing (TC) Program and is a publicly available dataset used in previous research. The TC Program aims to enhance visibility in modern computing systems, which have become increasingly opaque due to their complexity, making it difficult to detect attacks such as advanced persistent threats (APTs) \cite{Darpa-TC}.

In 2018, Engagement 3 (E3) of the TC Data was released, followed by Engagement 5 (E5) in 2019 \cite{Darpa-E3}\cite{Darpa-E5}. This experiment used three types of data: E3 Theia, E5 Theia, and E5 Marple.

\vskip.5\baselineskip
\noindent\textbf{Engagement 3 (E3) Theia}

This dataset consists of log data from a single Ubuntu 12.04 OS host, including attacks such as backdoor installation via Firefox vulnerabilities and phishing emails with malicious links. A total of 28.3 GB was used as evaluation data.

\vskip.5\baselineskip
\noindent\textbf{Engagement 5 (E5) Theia}

This dataset contains log data from three Ubuntu 12.04 OS hosts, including attacks such as backdoor installation via Firefox vulnerabilities and C2 server communications. A total of 295.8 GB was used as evaluation data.

\vskip.5\baselineskip
\noindent\textbf{Engagement 5 (E5) Marple}

This dataset consists of log data from a single Windows 7 OS host, including attacks such as backdoor installation via Firefox vulnerabilities and C2 server communications. A total of 120.7 GB was used as evaluation data.

\vskip.5\baselineskip
For each of the three datasets, portions of the log data that were different from the evaluation data were used as labeled data for extracting benign activities. The proportion of labeled data used for benign activity extraction relative to the evaluation data is shown in Table \ref{tab:datalist}.

\begin{table}[tb]
\caption{Data Used for Extracting Benign Activities}
\label{tab:datalist}
\centering
  \begin{tabular}{ccc}
   \hline
   Dataset & Data size & Proportion to Evaluation Data \\
   \hline \hline
   E3 Theia-A & 3.8GB & 13.4\% \\

   E3 Theia-B & 3.8GB & 13.4\% \\

   E3 Theia-C & 3.8GB & 13.4\% \\
   \hline
   E5 Theia-A & 4.0GB & 1.35\% \\

   E5 Theia-B & 4.0GB & 1.35\% \\

   E5 Theia-C & 4.0GB & 1.35\% \\
   \hline
   E5 Marple-A & 3.6GB & 2.98\% \\

   E5 Marple-B & 3.6GB & 2.98\% \\

   E5 Marple-C & 3.8GB & 3.15\% \\
   \hline
  \end{tabular}
\end{table}

\subsection{Dataset Analysis Results}

This dataset does not include annotations specifying which lines in the log data correspond to malicious activities. Therefore, we analyzed the dataset and created a list of nodes associated with attack activities within the log data. To identify attack activities, we parsed the dataset using command lines, filenames, and IP addresses provided in the Ground Truth documents by DARPA as Indicators of Compromise (IoCs) \cite{Darpa-E3GroundTruth}\cite{Darpa-E5GroundTruth}. 

\begin{figure}[tb]
\begin{center}
\includegraphics[scale=0.6]{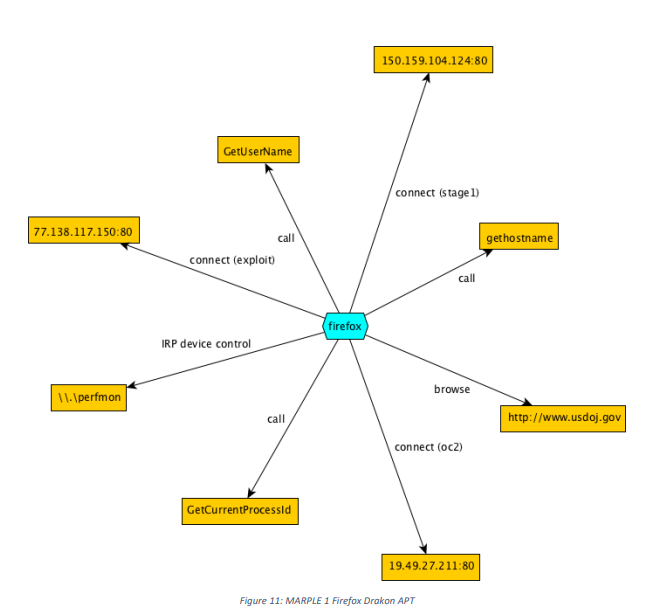}
\end{center}
\caption{Malicious Activities Based on E5 Marple Ground Truth\cite{Darpa-E5GroundTruth}}
\label{fig:marple-GroundTruth}
\end{figure}

All nodes containing the identified IoCs, as well as nodes related to events that included these IoCs, were classified as malicious nodes. For E5 Marple, the Ground Truth document provided a Provenance Graph illustrating attack activities, which was used as a reference to create the list of malicious nodes (Figure \ref{fig:marple-GroundTruth}). When analyzing the dataset, if an IoC mentioned in the document or a corresponding node in the Provenance Graph was not found, it was not included in the list of malicious nodes.

Additionally, to estimate the characteristics of each dataset, we extracted the total number of occurrences of each system call and command line used throughout the log data. For command lines, we obtained the top 10 most frequently occurring commands along with their proportion of the total command line occurrences.

Below, we present the number of analyzed malicious nodes and the estimated characteristics for each dataset.

\vskip.5\baselineskip
\noindent\textbf{Engagement 3 (E3) Theia}

A total of 81 malicious nodes were identified.

\begin{figure}[tb]
\begin{center}
\includegraphics[scale=0.5]{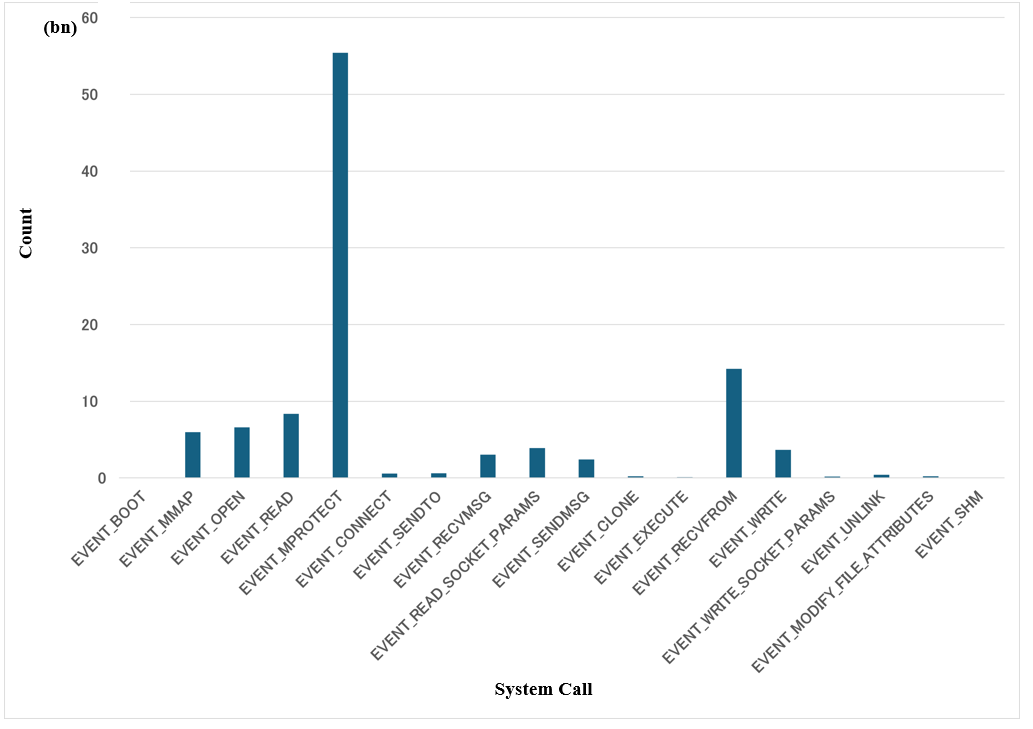}
\end{center}
\caption{E3 Theia Number of Events by System Call}
\label{fig:E3theia-syscall}
\end{figure}

Figure \ref{fig:E3theia-syscall} shows the number of occurrences of each system call in the E3 Theia dataset, while Table \ref{tab:E3theia-cmdline} presents the frequency of command line occurrences. The results indicate that MPROTECT and RECVFROM system calls appear relatively frequently. In terms of command line occurrence ratio, the top three commands account for more than half of the total command line executions. The command line data also revealed that applications such as Firefox and Thunderbird were frequently executed.

\begin{table}[tb]
\caption{E3 Theia Top 10 Most Frequent Command Lines}
\label{tab:E3theia-cmdline}
\centering
  \begin{tabular}{lrr}
   \hline
   Command line & Count & Proportion to All Command Lines \\
   \hline \hline
   /usr/bin/firefox & 77985 & 28.01\% \\
   \hline
    -bash & 53061 & 19.06\% \\
    \hline
    /bin/sh -e /usr/lib/update-notifier/ \\update-motd-updates-available & 25917 & 9.31\% \\
    \hline
    /usr/lib/postgresql/9.1/bin/postgres -D /var/\\lib/postgresql/9.1/main -c config\_file=/etc/\\postgresql /9.1/main/postgresql.conf & 10800 & 3.88\% \\
    \hline
    /usr/bin/python /usr/lib/\\unity-scope-video-remote/unity-scope-video-remote & 7004 & 2.52\% \\
    \hline
    N/A & 6544 & 2.35\% \\
    \hline
    /usr/bin/pulseaudio --start --log-target=syslog & 5655 & 2.03\% \\
    \hline
    /usr/lib/thunderbird/thunderbird & 5586 & 2.01\% \\
    \hline
    run-parts --lsbsysinit /etc/update-motd.d & 3915 & 1.41\% \\
    \hline
    /usr/sbin/console-kit-daemon --no-daemon & 3815 & 1.37\% \\

   \hline
  \end{tabular}
\end{table}

\vskip.5\baselineskip
\noindent\textbf{Engagement 5 (E5) Theia}

A total of 4 malicious nodes were identified.

\begin{figure}[tb]
\begin{center}
\includegraphics[scale=0.5]{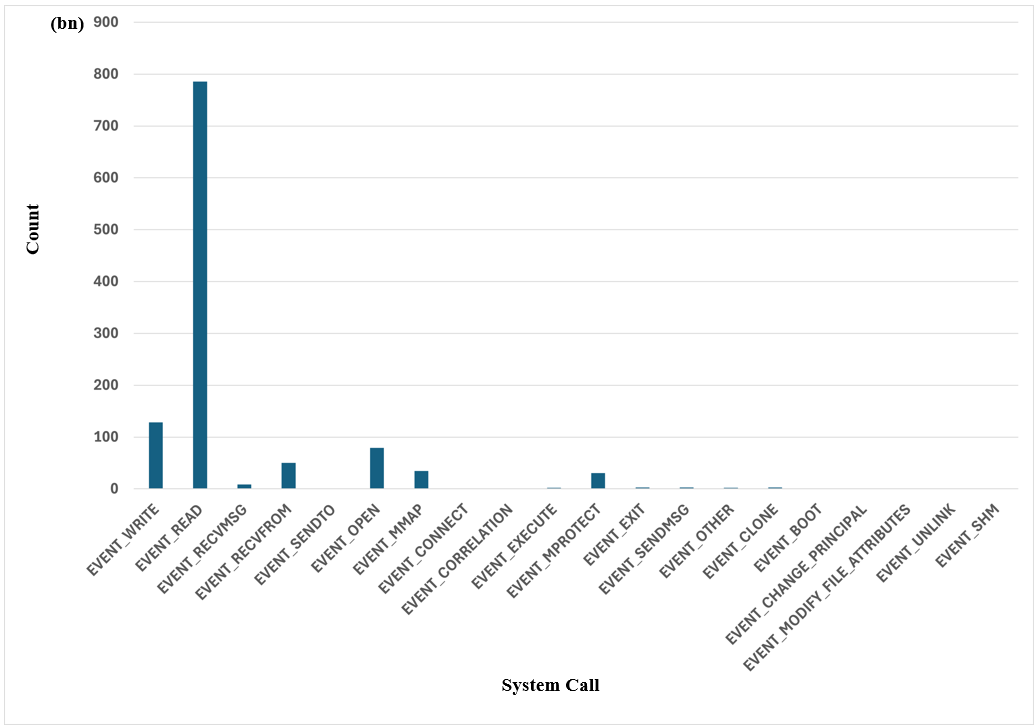}
\end{center}
\caption{E5 Theia Number of Events by System Call}
\label{fig:E5theia-syscall}
\end{figure}

Figure \ref{fig:E5theia-syscall} shows the number of occurrences of each system call in the E5 Theia dataset, while Table \ref{tab:E5theia-cmdline} presents the frequency of command line occurrences. The results indicate that READ and WRITE system calls appear significantly more frequently than others. In terms of command line occurrence ratio, the top three commands account for more than 70

\begin{table}[tb]
\caption{E5 Theia Top 10 Most Frequent Command Lines}
\label{tab:E5theia-cmdline}
\centering
  \begin{tabular}{lrr}
   \hline
   Command line & Count & Proportion to All Command Lines \\
   \hline\hline
    /bin/sh -e /usr/lib/update-notifier/\\update-motd-updates-available & 1259770 & 41.12\% \\
    \hline
    -bash & 866708 & 28.29\% \\
    \hline
    run-parts --lsbsysinit /etc/update-motd.d & 109023 & 3.56\% \\
    \hline
    /bin/sh -e /usr/lib/update-notifier/\\update-motd-hwe-eol & 84798 & 2.77\% \\
    \hline
    /sbin/modprobe -q \\-- net-pf-16-proto-16-family-nl80211 & 81572 & 2.66\% \\
    \hline
    /bin/sh /etc/update-motd.d/50-landscape-sysinfo & 80304 & 2.62\% \\
    \hline
    /bin/sh /usr/lib/update-notifier/\\update-motd-fsck-at-reboot & 60732 & 1.98\% \\
    \hline
    kworker/u:2 & 51472 & 1.68\% \\
    \hline
    /usr/bin/python /usr/bin/landscape-sysinfo & 39439 & 1.29\% \\
    \hline
    /bin/sh /etc/update-motd.d/00-header & 36344 & 1.19\% \\
   \hline
  \end{tabular}
\end{table}

\begin{figure}[tb]
\begin{center}
\includegraphics[scale=0.5]{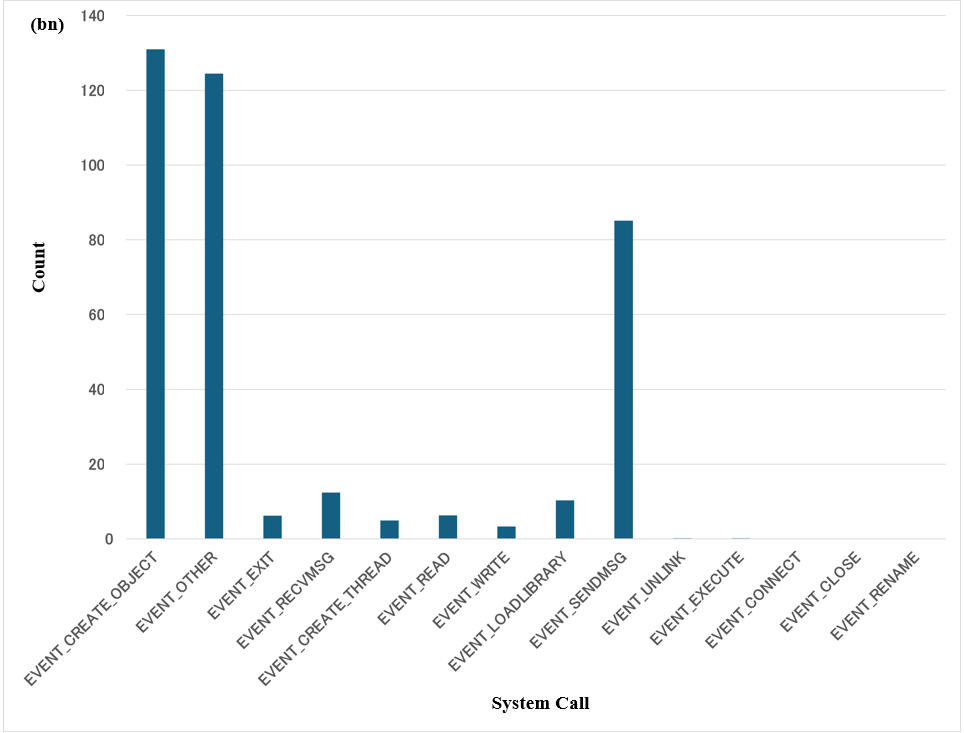}
\end{center}
\caption{E5 Marple Number of Events by System Call}
\label{fig:Marple-syscall}
\end{figure}

\vskip.5\baselineskip
\noindent\textbf{Engagement 5 (E5) Marple}

A total of 10 malicious nodes were identified.  
The attack activity diagram provided in the Ground Truth document (Figure \ref{fig:marple-GroundTruth}) consists of 9 nodes. However, when searching for corresponding node information, multiple matching nodes were found. In cases where it was difficult to determine which specific nodes were malicious based on attack timestamps and other contextual information, all such nodes were classified as malicious.

Figure \ref{fig:Marple-syscall} shows the number of occurrences of each system call in the E5 Marple dataset, while Table \ref{tab:E5Marple-cmdline} presents the frequency of command line occurrences. The results indicate that CREATE\_OBJECT, OTHER, and SENDMSG system calls appeared frequently. Compared to E3 Theia and E5 Theia, the distribution of command line occurrences in E5 Marple was more balanced, without significant biases.  

Table \ref{tab:E5Marple-cmdline-after10} presents the frequency of command lines beyond the top 10 most frequently occurring commands. The command line data revealed that operations frequently involved document files, with only the filenames differing between command executions.

\begin{table}[tb]
\caption{E5 Marple Top 10 Most Frequent Command Lines}
\label{tab:E5Marple-cmdline}
\centering
  \begin{tabular}{p{8cm}rr}
   \hline
   Command line & Count & Proportion to All Command Lines \\
   \hline\hline
    "C:\textbackslash Program Files\textbackslash OpenSSH-Win64\textbackslash ssh-agent.exe" 268 & 30269 & 2.06\% \\
    \hline
    "C:\textbackslash Program Files\textbackslash OpenSSH-Win64\textbackslash sshd.exe" & 13884 & 0.95\% \\
    \hline
    C:\textbackslash Windows\textbackslash system32\textbackslash DllHost.exe /Processid:{E10F6C3A-F1AE-4ADC-AA9D-2FE65525666E} & 11962 & 0.82\% \\
    \hline
    "C:\textbackslash Program Files\textbackslash OpenSSH-Win64\textbackslash ssh-shellhost.exe" & 11383 & 0.78\% \\
    \hline
    C:\textbackslash Windows\textbackslash system32\textbackslash cmd.exe & 11382 & 0.78\% \\
    \hline
    "C:\textbackslash Windows\textbackslash system32\textbackslash SearchFilterHost.exe" 0 512 516 524 65536 520 & 9396 & 0.64\% \\
    \hline
    "C:\textbackslash Program Files\textbackslash OpenSSH-Win64\textbackslash ssh-agent.exe" 272 & 8941 & 0.61\% \\
    \hline
    "C:\textbackslash Program Files\textbackslash OpenSSH-Win64\textbackslash ssh-agent.exe" 280 & 5408 & 0.37\% \\
    \hline
    "C:\textbackslash Windows\textbackslash system32\textbackslash SearchFilterHost.exe" 0 516 520 528 65536 524 & 2970 & 0.20\% \\
    \hline
    systeminfo & 2693 & 0.18\% \\
    \hline
    
  \end{tabular}
\end{table}

\begin{table}[tb]
\caption{E5 Marple Command Line Counts (Ranks 11–20)}
\label{tab:E5Marple-cmdline-after10}
\centering
  \begin{tabular}{p{8cm}rr}
   \hline
   Command line & Count & Proportion to All Command Lines \\
   \hline\hline
    netstat -na & 1038 & 0.07\% \\
    \hline
    C:\textbackslash Windows\textbackslash system32\textbackslash DllHost.exe /Processid:{F9717507-6651-4EDB-BFF7-AE615179BCCF} & 876 & 0.06\% \\
    \hline
    hostname & 744 & 0.05\% \\
    \hline
    ipconfig & 723 & 0.05\% \\
    \hline
    tasklist & 713 & 0.05\% \\
    \hline
    C:\textbackslash Windows\textbackslash system32\textbackslash cmd.exe /c Python "C:\textbackslash Users\textbackslash darpa\textbackslash ...\textbackslash oletoolsmraptor3.py" "C:\textbackslash Users\textbackslash admin\textbackslash Documents\textbackslash Documents\textbackslash -guide.com.docx" >temp.txt & 711 & 0.05\% \\
    \hline
    C:\textbackslash Windows\textbackslash system32\textbackslash cmd.exe /c Python "C:\textbackslash Users\textbackslash darpa\textbackslash ...\textbackslash oletoolsmraptor3.py" "C:\textbackslash Users\textbackslash admin\textbackslash Documents\textbackslash Documents\textbackslash ... redds.docx" >temp.txt & 692 & 0.05\% \\
    \hline
    C:\textbackslash Windows\textbackslash system32\textbackslash cmd.exe /c Python "C:\textbackslash Users\textbackslash darpa\textbackslash ...\textbackslash oletoolsmraptor3.py" "C:\textbackslash Users\textbackslash admin\textbackslash Documents\textbackslash Documents\textbackslash ... limns.docx" >temp.txt & 690 & 0.05\% \\
    \hline
    C:\textbackslash Windows\textbackslash system32\textbackslash cmd.exe /c Python "C:\textbackslash Users\textbackslash darpa\textbackslash ...\textbackslash oletoolsmraptor3.py" "C:\textbackslash Users\textbackslash admin\textbackslash Documents\textbackslash Documents\textbackslash \\admissibility.docx" >temp.txt & 685 & 0.05\% \\
    \hline
    C:\textbackslash Windows\textbackslash system32\textbackslash cmd.exe /c Python "C:\textbackslash Users\textbackslash darpa\textbackslash ...\textbackslash oletoolsmraptor3.py" "C:\textbackslash Users\textbackslash admin\textbackslash Documents\textbackslash Documents\textbackslash \\aefaldy.docx" >temp.txt & 685 & 0.05\% \\
    \hline
    
  \end{tabular}
\end{table}

\clearpage
\section{Experimental Results}

Table \ref{tab:label-count} presents the number of labels generated for each labeled dataset and the execution time required from data preprocessing to labeling. The number of labels was relatively high when using E3 Theia, reaching a maximum of 16,010 labels. The lowest number of labels was obtained when using E5 Marple-B, with 884 labels generated.

\begin{figure}[tb]
\begin{center}
\includegraphics[scale=0.8]{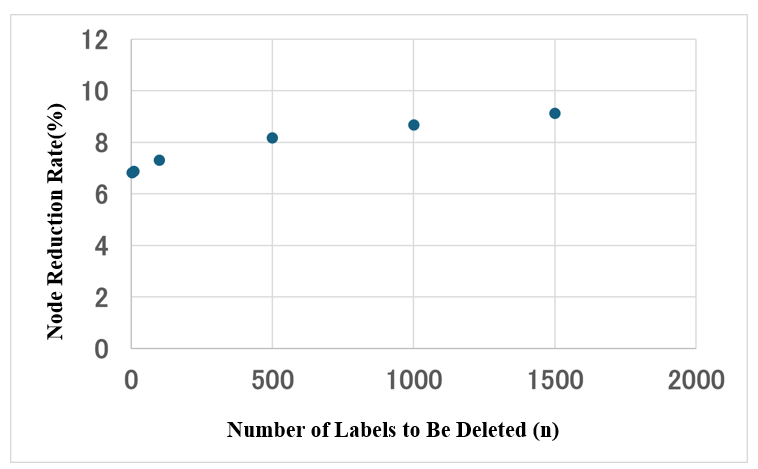}
\end{center}
\caption{Change in Node Reduction Rate by Number of Labels to Be Deleted(E3 Theia-A)}
\label{fig:E3theiaA}
\end{figure}

\begin{figure}[tb]
\begin{center}
\includegraphics[scale=0.8]{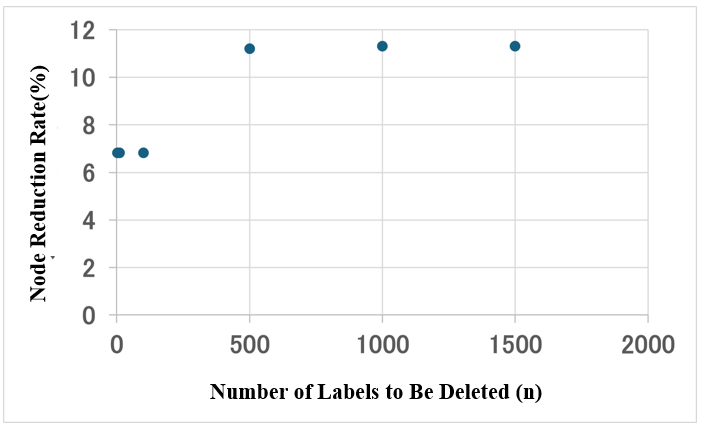}
\end{center}
\caption{Change in Node Reduction Rate by Number of Labels to Be Deleted(E3 Theia-B)}
\label{fig:E3theiaB}
\end{figure}

\begin{figure}[tb]
\begin{center}
\includegraphics[scale=0.8]{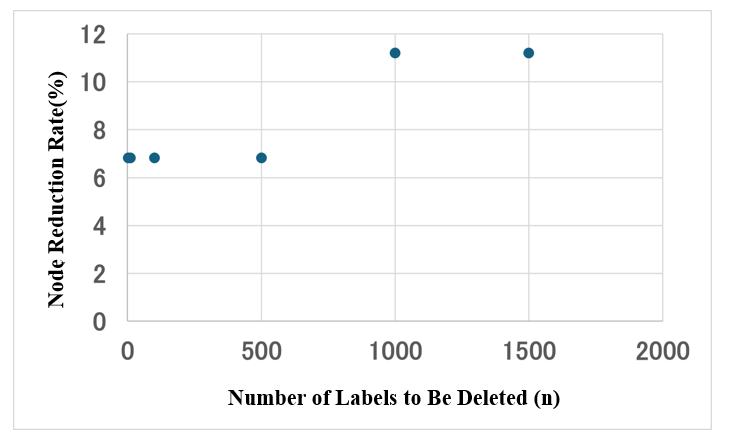}
\end{center}
\caption{Change in Node Reduction Rate by Number of Labels to Be Deleted(E3 Theia-C)}
\label{fig:E3theiaC}
\end{figure}

Tables \ref{tab:E3Theia-A}, \ref{tab:E3Theia-B}, and \ref{tab:E3Theia-C} present the experimental results using E3 Theia. Additionally, Figures \ref{fig:E3theiaA}, \ref{fig:E3theiaB}, and \ref{fig:E3theiaC} illustrate the change in node reduction rate based on the number of labels removed. The total number of nodes in the dependency graph before reduction, after applying data preprocessing as described in Section \ref{preprocess}, was 54,522 nodes. The node reduction rate is defined as the proportion of removed nodes relative to the total number of nodes before reduction. The node reduction rate remained similar across the three labeled datasets.

\begin{figure}[tb]
\begin{center}
\includegraphics[scale=0.8]{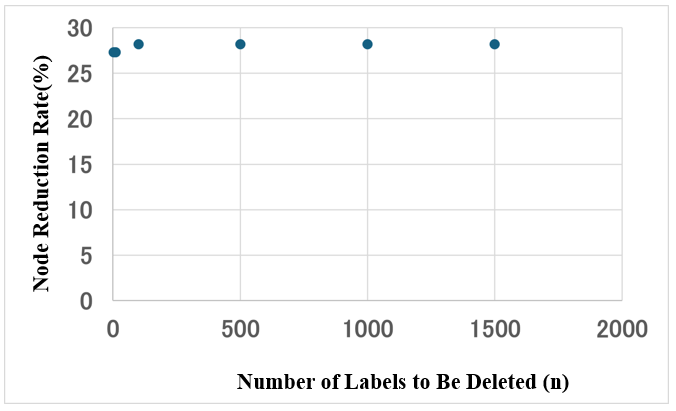}
\end{center}
\caption{Change in Node Reduction Rate by Number of Labels to Be Deleted(E5 Theia-A)}
\label{fig:E5theiaA}
\end{figure}

\begin{figure}[tb]
\begin{center}
\includegraphics[scale=0.8]{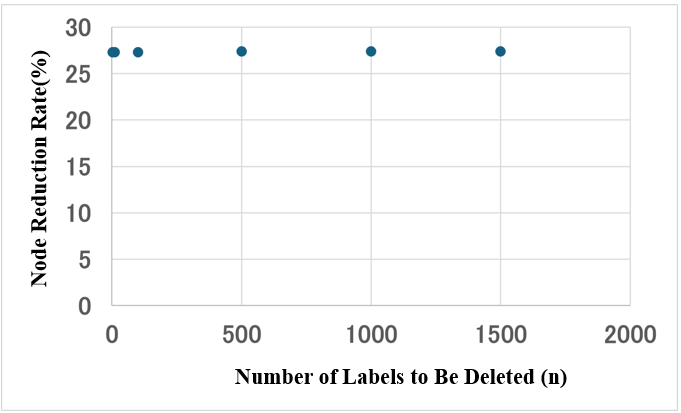}
\end{center}
\caption{Change in Node Reduction Rate by Number of Labels to Be Deleted(E5 Theia-B)}
\label{fig:E5theiaB}
\end{figure}

\begin{figure}[tb]
\begin{center}
\includegraphics[scale=0.8]{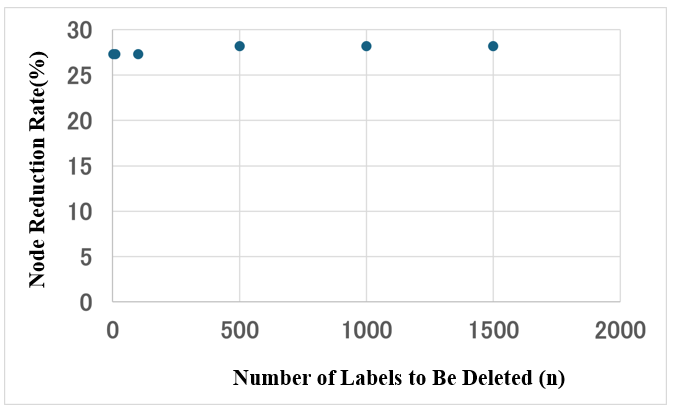}
\end{center}
\caption{Change in Node Reduction Rate by Number of Labels to Be Deleted(E5 Theia-C)}
\label{fig:E5theiaC}
\end{figure}

Tables \ref{tab:E5Theia-A}, \ref{tab:E5Theia-B}, and \ref{tab:E5Theia-C} present the experimental results using E5 Theia. Additionally, Figures \ref{fig:E5theiaA}, \ref{fig:E5theiaB}, and \ref{fig:E5theiaC} illustrate the change in node reduction rate based on the number of labels removed. The total number of nodes in the dependency graph before reduction for E5 Theia was 3,034,111 nodes. Similar to E3 Theia, the node reduction rate remained consistent across the three labeled datasets.

\begin{figure}[tb]
\begin{center}
\includegraphics[scale=0.8]{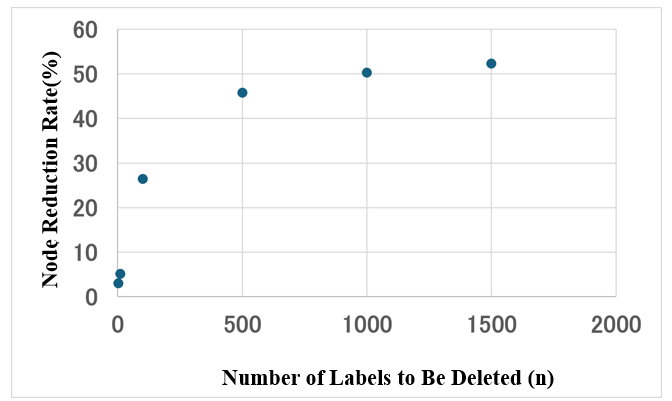}
\end{center}
\caption{Change in Node Reduction Rate by Number of Labels to Be Deleted(E5 Marple-A)}
\label{fig:E5marpleA}
\end{figure}

\begin{figure}[tb]
\begin{center}
\includegraphics[scale=0.8]{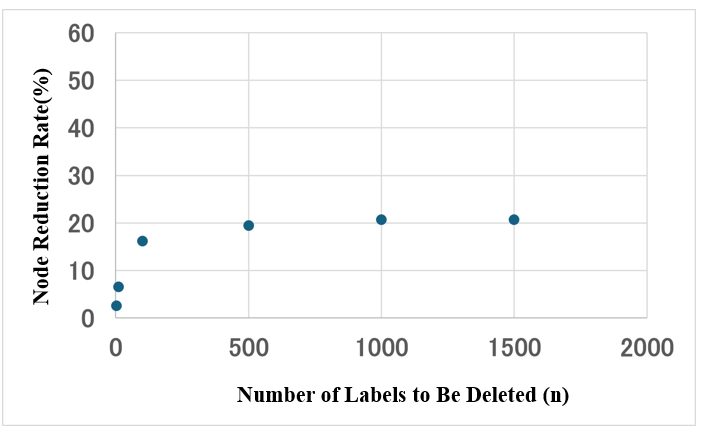}
\end{center}
\caption{Change in Node Reduction Rate by Number of Labels to Be Deleted(E5 Marple-B)}
\label{fig:E5marpleB}
\end{figure}

\begin{figure}[tb]
\begin{center}
\includegraphics[scale=0.8]{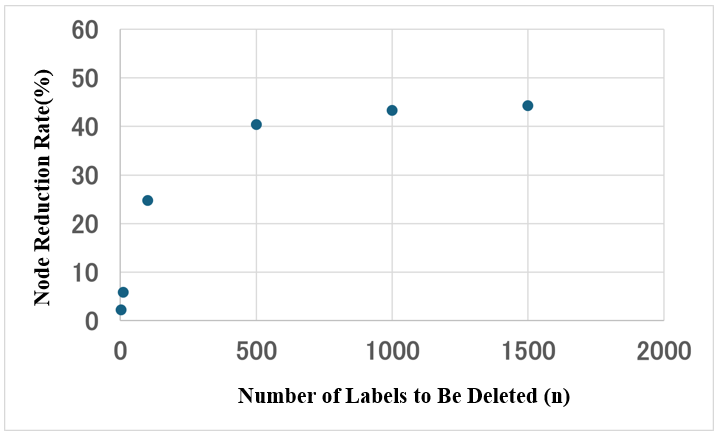}
\end{center}
\caption{Change in Node Reduction Rate by Number of Labels to Be Deleted(E5 Marple-C)}
\label{fig:E5marpleC}
\end{figure}

Tables \ref{tab:E5Marple-A}, \ref{tab:E5Marple-B}, and \ref{tab:E5Marple-C} present the experimental results using E5 Marple. Additionally, Figures \ref{fig:E5marpleA}, \ref{fig:E5marpleB}, and \ref{fig:E5marpleC} illustrate the change in node reduction rate based on the number of labels removed. The total number of nodes in the dependency graph before reduction for E5 Marple was 12,172,296 nodes. When using E5 Marple-B, the node reduction rate showed a tendency to increase less significantly compared to the other two datasets, even as the number of removed labels increased.

Within the scope of this experiment, no False Negatives (FN) occurred regardless of the labeled dataset used. As a result, the False Positives (FP) can be calculated by subtracting the number of malicious nodes from the total number of remaining nodes.  

Table \ref{tab:average} presents the average reduction rate and average execution time for each dataset when using the three labeled datasets. The minimum and maximum average reduction rates were both observed when using E5 Marple, ranging from 2.62\% to 39.1\%. The execution time tended to increase as the average reduction rate increased.

\begin{table}[b]
\caption{Number of Generated Labels and Execution Time for Labeling}
\label{tab:label-count}
\centering
  \begin{tabular}{crr}
   \hline
   Dataset & Label Count & Execution Time(sec) \\
   \hline \hline
   E3 Theia-A & 16,010 & 1,163 \\

   E3 Theia-B & 12,531 & 616.7 \\

   E3 Theia-C & 11,839 & 607.0 \\
   \hline
   E5 Theia-A & 2,879 & 168.3\\

   E5 Theia-B & 2,384 & 159.6 \\

   E5 Theia-C & 3,557 & 172.8 \\
   \hline
   E5 Marple-A & 6,772 & 1,913\\

   E5 Marple-B & 884 & 1,112\\

   E5 Marple-C & 3,658 & 945.9 \\
   \hline
  \end{tabular}
\end{table}

\begin{table}[tb]
\centering
\caption{Experimental Results(E3 Theia-A)}
\label{tab:E3Theia-A}
\begin{tabular}{rrcrrr}
\hline
Number of Labels \\ to Be Deleted (n) &Node Count&FN&FP&Node Reduction Rate(\%)&Execution Time(sec)\\
\hline  
\hline 
3&50,802&0&50,721&6.82&7.264\\
10&50,778&0&50,697&6.87&8.701\\
100&50,541&0&50,460&7.30&16.09\\
500&50,064&0&49,983&8.17&44.08\\
1000&49,796&0&49,715&8.67&77.24\\
1500&49,549&0&49,468&9.12&103.1\\
\hline
\end{tabular}
\end{table}

\clearpage

\begin{table}[tb]
\centering
\caption{Experimental Results(E3 Theia-B)}
\label{tab:E3Theia-B}
\begin{tabular}{rrcrrr}
\hline
Number of Labels \\ to Be Deleted (n) &Node Count&FN&FP&Node Reduction Rate(\%)&Execution Time(sec)\\
\hline  
\hline 
3&50,802&0&50,721&6.82&7.396\\
10&50,802&0&50,721&6.82&8.682\\
100&50,798&0&50,717&6.83&15.96\\
500&48,393&0&48,312&11.2&42.88\\
1000&48,383&0&48,302&11.3&68.49\\
1500&48,380&0&48,299&11.3&90.37\\
\hline
\end{tabular}
\end{table}

\clearpage

\begin{table}[tb]
\centering
\caption{Experimental Results(E3 Theia-C)}
\label{tab:E3Theia-C}
\begin{tabular}{rrcrrr}
\hline
Number of Labels \\ to Be Deleted (n) &Node Count&FN&FP&Node Reduction Rate(\%)&Execution Time(sec)\\
\hline  
\hline 
3&50,802&0&50,721&6.82&6.83\\
10&50,802&0&50,721&6.82&7.779\\
100&50,801&0&50,720&6.82&15.98\\
500&50,800&0&50,719&6.83&43.89\\
1000&48,398&0&48,317&11.2&69.63\\
1500&48,398&0&48,317&11.2&91.43\\
\hline
\end{tabular}
\end{table}

\clearpage

\begin{table}[tb]
\centering
\caption{Experimental Results(E5 Theia-A)}
\label{tab:E5Theia-A}
\begin{tabular}{rrcrrr}
\hline
Number of Labels \\ to Be Deleted (n) &Node Count&FN&FP&Node Reduction Rate(\%)&Execution Time(sec)\\
\hline  
\hline 
3&2,206,930&0&2,206,926&27.3&795.4\\
10&2,206,813&0&2,206,809&27.3&829.1\\
100&2,178,684&0&2,178,680&28.2&1100\\
500&2,177,642&0&2,177,638&28.2&1906\\
1000&2,177,392&0&2,177,388&28.2&2638\\
1500&2,177,329&0&2,177,325&28.2&3136\\
\hline
\end{tabular}
\end{table}

\clearpage

\begin{table}[tb]
\centering
\caption{Experimental Results(E5 Theia-B)}
\label{tab:E5Theia-B}
\begin{tabular}{rrcrrr}
\hline
Number of Labels \\ to Be Deleted (n) &Node Count&FN&FP&Node Reduction Rate(\%)&Execution Time(sec)\\
\hline  
\hline 
3&2,206,909&0&2,206,905&27.3&783.9\\
10&2,206,769&0&2,206,765&27.3&810.7\\
100&2,205,545&0&2,205,541&27.3&1051\\
500&2,204,039&0&2,204,035&27.4&1794\\
1000&2,203,720&0&2,203,716&27.4&2435\\
1500&2,203,584&0&2,203,580&27.4&2879\\
\hline
\end{tabular}
\end{table}

\clearpage

\begin{table}[tb]
\centering
\caption{Experimental Results(E5 Theia-C)}
\label{tab:E5Theia-C}
\begin{tabular}{rrcrrr}
\hline
Number of Labels \\ to Be Deleted (n) &Node Count&FN&FP&Node Reduction Rate(\%)&Execution Time(sec)\\
\hline  
\hline 
3&2,206,900&0&2,206,896&27.3&797.8\\
10&2,206,790&0&2,206,786&27.3&822.7\\
100&2,205,921&0&2,205,917&27.3&1,085\\
500&2,177,493&0&2,177,489&28.2&1,932\\
1000&2,177,258&0&2,177,254&28.2&2,654\\
1500&2,177,248&0&2,177,244&28.2&3,169\\
\hline
\end{tabular}
\end{table}

\clearpage

\begin{table}[tb]
\centering
\caption{Experimental Results(E5 Marple-A)}
\label{tab:E5Marple-A}
\begin{tabular}{rrcrrr}
\hline
Number of Labels \\ to Be Deleted (n) &Node Count&FN&FP&Node Reduction Rate(\%)&Execution Time(sec)\\
\hline  
\hline 
3&11,803,667&0&11,803,657&3.03&558.7\\
10&11,545,745&0&11,545,735&5.15&717.1\\
100&8,956,381&0&8,956,371&26.4&2,091\\
500&6,600,627&0&6,600,617&45.8&5,768\\
1000&6,048,300&0&6,048,290&50.3&9,130\\
1500&5,800,663&0&5,800,653&52.3&11,690\\
\hline
\end{tabular}
\end{table}

\clearpage

\begin{table}[tb]
\centering
\caption{Experimental Results(E5 Marple-B)}
\label{tab:E5Marple-B}
\begin{tabular}{rrcrrr}
\hline
Number of Labels \\ to Be Deleted (n) &Node Count&FN&FP&Node Reduction Rate(\%)&Execution Time(sec)\\
\hline  
\hline 
3&11,855,130&0&11,855,120&2.61&527.5\\
10&11,373,838&0&11,373,828&6.56&627.8\\
100&9,648,817&0&9,648,807&16.2&1,516\\
500&9,795,840&0&9,795,830&19.5&3,008\\
1000&9,648,817&0&9,648,807&20.7&3,603\\
1500&9,648,817&0&9,648,807&20.7&3,631\\
\hline
\end{tabular}
\end{table}

\clearpage

\begin{table}[tb]
\centering
\caption{Experimental Results(E5 Marple-C)}
\label{tab:E5Marple-C}
\begin{tabular}{rrcrrr}
\hline
Number of Labels \\ to Be Deleted (n) &Node Count&FN&FP&Node Reduction Rate(\%)&Execution Time(sec)\\
\hline  
\hline 
3&11,903,443&0&11,903,433&2.21&538.9\\
10&11,459,815&0&11,459,805&5.85&678.7\\
100&9,154,142&0&9,154,132&24.8&1886\\
500&7,258,716&0&7,258,706&40.4&5265\\
1000&6,900,864&0&6,900,854&43.3&8375\\
1500&6,779,474&0&6,779,464&44.3&10370\\
\hline
\end{tabular}
\end{table}

\clearpage

\clearpage

\begin{table}[tb]
    \centering
    \caption{Average Reduction Rate and Average Execution Time for Each Dataset}
    \label{tab:average}
    \renewcommand{\arraystretch}{1.5}
    \begin{tabular}{l|cc|cc}
        \toprule
        \textbf{Dataset} & \multicolumn{2}{c|}{\textbf{Average Reduction Rate (\%)}} & \multicolumn{2}{c}{\textbf{Average Execution Time (sec)}} \\
        & Min ($n=3$) & Max ($n=1500$) & Min ($n=3$) & Max ($n=1500$) \\
        \hline \hline 
        E3 Theia & 6.82 & 10.5 & 7.163 & 94.97 \\
        E5 Theia & 27.3 & 27.9 & 792.4 & 3,061 \\
        E5 Marple & 2.62 & 39.1 & 541.7 & 8,563 \\
        \bottomrule
    \end{tabular}
    
\end{table}

%% file: Chapter5/chapter5.tex
\chapter{Discussion}

\section{Effectiveness of Dependency Reduction}

The experimental results demonstrated that extracting benign activities and removing them from the dependency graph effectively reduces dependency relationships. Within the scope of this experiment, no False Negatives (FN) occurred, suggesting that node sets containing malicious nodes exhibited distinct characteristics from those with benign labels. 

Regarding the node reduction rate, Table \ref{tab:E5Marple-A} shows that a maximum reduction of 52.3\% was achieved. Furthermore, based on the average reduction rates presented in Table \ref{tab:average}, it is estimated that approximately 6.8\% to 39\% of system activities can be identified as frequently occurring benign activities.

In addition, experiments were conducted using three different labeled datasets per dataset. The experimental results shown in Tables \ref{tab:E3Theia-A}, \ref{tab:E3Theia-B}, and \ref{tab:E3Theia-C} for E3 Theia, and Tables \ref{tab:E5Theia-A}, \ref{tab:E5Theia-B}, and \ref{tab:E5Theia-C} for E5 Theia, indicated that changing the labeled dataset did not significantly affect the reduction rate. This suggests that the proposed method effectively extracted frequently occurring benign activities within the system. Since frequently occurring benign activities are not limited to a specific dataset but likely appear across different periods in the same host or environment, the method appears robust.

On the other hand, Table \ref{tab:E5Marple-B} shows that E5 Marple-B exhibited a lower node reduction rate compared to the results obtained with E5 Marple-A and E5 Marple-C. This is likely due to the log structure characteristics of TC Data. As illustrated in Figure \ref{fig:log-graph}, TC Data logs first record process, file, and network flow information to define node information, followed by event logs that involve the corresponding nodes. Since node and event information are recorded separately, if the labeled dataset contains many event logs referencing node information defined earlier in the logs, the proposed method may fail to retrieve the necessary node information. 

Additionally, Table \ref{tab:label-count} shows that E5 Marple-B generated fewer labels than other datasets. This suggests that a significant portion of required node information was missing in E5 Marple-B, which may have contributed to the lower reduction rate compared to other E5 Marple experiments.

\section{Dataset Characteristics and Benign Activity Proportions}

Within the scope of this experiment, Table \ref{tab:average} shows that E5 Marple exhibited the highest average node reduction rate (39.1\%) when considering the top 1500 labels for removal. Conversely, the lowest reduction rate for the top 1500 labels was observed in E3 Theia. However, when considering only the top 3 labels, E5 Marple exhibited the lowest reduction rate. Additionally, E5 Theia’s average reduction rate remained nearly unchanged regardless of whether the number of removed labels was 3 or 1500.

These results indicate that dataset characteristics significantly influenced the node reduction rate.

- E3 Theia (Table \ref{tab:E3theia-cmdline}):  
  The most frequently occurring command lines included applications such as Firefox and Thunderbird, suggesting that this dataset was collected from a general-purpose computing environment.

- E5 Theia (Table \ref{tab:E5theia-cmdline}):  
  System management and update processes accounted for a large portion of command executions, with the top commands covering most of the total occurrences. Additionally, Figure \ref{fig:E5theia-syscall} shows that READ and WRITE system calls were dominant, indicating a server-like system with repetitive operations.

Since E5 Theia exhibited a highly skewed distribution of command occurrences, frequently occurring activities were highly repetitive. Consequently, removing only the top 3 frequent activities eliminated most benign activities, causing little change in reduction rate even when additional labels were removed. Furthermore, since E5 Theia exhibited a higher reduction rate than E3 Theia, the method is likely to be more effective for systems with repetitive specialized operations than for general-purpose systems.

- E5 Marple (Table \ref{tab:E5Marple-cmdline}):  
  The command distribution was more balanced compared to other datasets. Additionally, frequent document file processing operations were observed, as shown in Table \ref{tab:E5Marple-cmdline-after10}. Although the overall command occurrence distribution appeared balanced, these operations were highly repetitive, suggesting that E5 Marple represents a specialized system primarily handling document file analysis. This likely contributed to the high node reduction rate.

Additionally, as shown in Table \ref{tab:datalist}, although E3 Theia contained a larger proportion of evaluation data than the other datasets, E5 Theia and E5 Marple exhibited higher reduction rates. This suggests that a large volume of data is not necessarily required for extracting benign activities, and that a dataset covering approximately 1.35\% to 3.15\% of the reduction target may be sufficient. Since frequently occurring benign activities are naturally present in even small log datasets, the method can effectively extract them.

\section{Limitations and Challenges of the Proposed Method}

\subsection{Limitations of the Proposed Method}

The proposed method assumes prior knowledge of all malicious nodes in the dataset. As a result, it is currently unable to analyze log data in real-time to extract benign activities. Instead, the method can be utilized by preprocessing collected logs from a specific environment, identifying malicious nodes within that dataset, and storing extracted benign activities to reduce the search space when analyzing other logs.

Additionally, the method relies on natural language data (e.g., command lines), which differ across operating systems. Therefore, handling log data from multiple OS environments may pose classification challenges.

\subsection{Potential for False Negatives (FN)}

Although no FN occurred in this experiment, the possibility of FN remains.

- Malicious nodes involved in both benign and malicious events were classified as malicious nodes.  
- If a benign node shares identical attributes (filename, command line, or network flow) with a malicious node, it may lead to FN if the labeling order results in incorrect classification.  
- This issue could be mitigated by defining benign activities based on edges rather than nodes, as nodes may participate in both benign and malicious events.

\subsection{Adaptation to Other Data Formats}

The proposed method is currently tailored for TC Data logs. To generalize it for other log formats, the system must be modified to flexibly extract provenance and node information. Additionally, even for TC Data, experiments revealed that certain labeled datasets lacked necessary node information, requiring an independent mechanism to retain node information across the entire dataset.

\subsection{Optimization of Parameters}

The method includes parameters such as node set size and cosine similarity threshold, which can influence results. In this experiment, node set size was fixed at 5, but further validation is needed to determine the optimal size. The cosine similarity threshold was set to 1.0, meaning only identical node sets were removed. Lowering this threshold could enable broader graph reduction, but initial tests showed minimal impact. This may be due to the curse of dimensionality in NODLINK’s 256-dimensional vector representation, necessitating further investigation into alternative similarity measures and NLP models.

\subsection{Application to Intrusion Detection Systems}

The extracted benign activities could potentially be applied to whitelist-based intrusion detection systems (IDS). Instead of whitelisting individual events or nodes, the method could whitelist activity patterns, leading to stricter security rules.

Additionally, in operational technology (OT) environments where systems perform highly repetitive tasks, whitelist-based IDS solutions are effective. Given the high effectiveness of our method on repetitive tasks, extending it to OT environments and other log formats could be a promising research direction.

%% file: Chapter6/chapter6.tex
\chapter{Related Work}

To address the dependency explosion problem, previous studies have proposed two main approaches: (1) excluding logs of individual events that do not affect analysis, and (2) applying weighting to edges and nodes to identify malicious activities.

\section{Approach: Excluding Individual Event Logs}

One approach involves removing logs that are not useful for attack analysis, thereby reducing the volume of logs to be analyzed. LogGC, proposed by Lee et al., defines events that do not impact subsequent system behavior as dead-end events and selectively retains only events related to process or file deletion for analysis \cite{LogGC}.

Furthermore, LogGC found that most deleted files were temporary files. Temporary files are defined as files that exist only within the lifetime of a single process. Since temporary files have a short lifespan and limited impact on the overall system, deleting temporary file removal events from the analysis further reduces the number of logs considered.

Another log reduction approach is CPR, proposed by Xu et al., which merges redundant events \cite{CPR}.  
For example, if a node A is read by node B twice, and no other events involving A or B occur in between, then the two read events can be considered equivalent, and one can be excluded from analysis.

Although these methods remove certain events such as temporary file deletions and redundant operations, their reduction capability is limited. Additionally, in environments where temporary file deletions are necessary for analysis, applying these rules may be infeasible.

\section{Approach: Identifying Malicious Activities}

To detect malicious activities, NODOZE, proposed by Hassan et al., assigns weights to all events along the causal dependency path of an alert and classifies them as normal or anomalous based on a threshold \cite{NODOZE}. 

NODOZE detects unusual system behavior by calculating transition frequencies between source and destination events. If a source event frequently transitions to a different destination than usual, it is considered anomalous. The anomaly score is also influenced by data flow volume over time—a process sending or receiving an unusual amount of data within a short period raises its anomaly score. Additionally:
- Executables and known malicious file extensions increase the anomaly score.
- Temporary files lower the anomaly score.

By filtering events based on their anomaly scores, only highly relevant attack dependencies remain in the graph.

Alternative methods use different weighting strategies. DEPIMPACT, proposed by Fang et al., assigns weights based on data flow volume, event timing, and node degree \cite{DEPIMPACT}. Events that exhibit:
1. Similar data flow characteristics to known malicious detection points,
2. Temporal proximity to a detected attack, or
3. A high ratio of outgoing to incoming connections,  
are classified as highly suspicious.

Additionally, Li et al. proposed NODLINK, a provenance-based EDR system \cite{nodlink}.  
NODLINK tracks event chains using Data Provenance and raises alerts at an activity level rather than an individual event level. To identify malicious activities, NODLINK, like DEPIMPACT and NODOZE, applies weighting based on:
- Anomaly scores,
- Distance from the detection point, and
- Inbound and outbound data flow proportions.

\section{Limitations of Malicious Activity Detection Approaches}

Methods that identify malicious activities often require:

\begin{enumerate}
\item Relearning new attack patterns as attacks evolve, and
\item Relearning benign behavior due to new server connections or system updates.
\end{enumerate}

These limitations highlight the need for an approach that reduces dependency explosion without frequent retraining.

%% file: Chapter7/chapter7.tex
\chapter{Conclusion}
This chapter summarizes the findings of this study and discusses future challenges.

\section{Summary}
Preventing initial intrusions by attackers entirely is challenging, making it increasingly important to establish a framework for attack analysis in anticipation of intrusions. As manual analysis by security experts becomes increasingly difficult, efforts are being made to simplify attack analysis through Data Provenance-based tracking and visualization of malicious activities. However, when linking a series of activities performed by attackers within a system, dependency explosion remains a problem, as routine system operations are included in the dependency graph, causing it to grow excessively large.

To address dependency explosion, existing methods have proposed excluding non-critical individual events or applying weighting to edges and nodes to identify malicious activities. However, no prior studies, to the best of our knowledge, have focused on extracting benign activities as units and utilizing them to reduce dependencies. Therefore, this study proposed a method to reduce dependencies at the level of benign activities by leveraging natural language processing and similarity-based classification. 

Using benign activities for dependency reduction is, as far as we know, the first attempt of its kind. This approach enables organizations to preserve their unique benign activities and establish rules for dependency reduction without requiring frequent retraining.

For evaluation, quantitative experiments were conducted using the publicly available TC Data dataset. The results demonstrated that frequently occurring benign activities can be extracted from log data and effectively used for dependency reduction. The findings suggest that approximately 6.8\% to 39\% of system activities can be classified as frequently occurring benign activities. Additionally, the amount of labeled log data required for training was found to be small, approximately 1.4\% of the total analysis dataset, indicating that a relatively small dataset is sufficient. Moreover, the proposed method was shown to be particularly effective for server environments and other systems that perform repetitive operations.

\section{Future Challenges}

\noindent\textbf{Edge-Based Classification}

The current node-based classification method leaves the possibility of False Negatives (FN). To improve accuracy, a classification method based on edges should be developed, where features are extracted from edges instead of nodes to more precisely distinguish between benign and malicious activities.

\vskip.5\baselineskip
\noindent\textbf{Expanding Data Coverage and Supporting Different Log Formats}

This study used only three datasets. To better understand the effectiveness of the method across different data characteristics, further validation with a wider variety of datasets is needed. Additionally, the proposed method is currently tailored for TC Data logs. Future improvements should focus on making the method more flexible and adaptable to different log formats.

\vskip.5\baselineskip
\noindent\textbf{Optimization of Parameters}

Several parameters, such as node set size and cosine similarity threshold, remain unoptimized in this study. Future work should explore lowering the similarity threshold to allow for more abstract definitions of benign activities and determine the optimal node set size for defining benign activities.

\vskip.5\baselineskip
\noindent\textbf{Application to Intrusion Detection Systems}

The extracted benign activities could be registered as a whitelist and used in an activity-based whitelist intrusion detection system. Since frequently occurring benign activities were found to be effective for repetitive systems, this approach may be particularly suitable for OT (Operational Technology) environments and ICS (Industrial Control Systems). Future research should explore adapting the method for diverse datasets and security environments.

%% file: acknowledgements.tex
\thispagestyle{empty}

\begin{center}
    {\LARGE\bf Acknowledgements}
\end{center}

I would like to express my sincere gratitude to Professor Ariyasu Suzaki of the Institute of Information Security, as well as Associate Professor Masaki Hashimoto of Kagawa University, for their invaluable guidance throughout this research. I am also deeply grateful to Professor Hiroshi Doi and Associate Professor Midori Inaba for their insightful advice and support.

Additionally, I extend my heartfelt appreciation to Dr. Shaofei Li (Ph.D. candidate, School of Computer Science, Peking University) for providing valuable advice and information sharing, as well as to the members of the Suzaki Laboratory and the Institute of Information Security for their continuous support and collaboration.

%% file: References/ref.tex
\begin{flushleft}

\end{flushleft}

%% file: dissertation.bbl
\begin{thebibliography}{99}
\bibitem{TrendMicro}
TREND MICRO. 2023 ANNUAL CYBERSECURITY REPORT. [Online]. https://www.trendmicro.com/vinfo/tw/security/research-and-analysis/threat-reports/roundup/calibrating-expansion-2023-annual-cybersecurity-threat-report. (visited on 2025-01-06)

\bibitem{APT}
Lockheed Martin. Cyber kill chain. [Online].https://www.lockheedmartin.com/en-us/capabilities/cyber/cyber-kill-chain.html. (visited on 2025-01-06)

\bibitem{MITRE}
MITRE. MITRE ATT\&CK. [Online]. https://attack.mitre.org/. (visited on 2025-01-06)

\bibitem{threat_report}
CrowdStrike Holdings, Inc. CrowdStrike 2024 Global Threat Report. [Online]. https://www.crowdstrike.com/global-threat-report/?utm\_campaign=globalthreatreport\&utm\_content=executivesummary. (visited on 2025-01-06)

\bibitem{SOC}
Vermeer, Mathew.  "Alert Alchemy: SOC Workflows and Decisions in the Management of NIDS Rules" Proceedings of the 2023 ACM SIGSAC Conference on Computer and Communications Security. 2023

\bibitem{workforse_report}
ISC2. How the Economy, Skills Gap and Artificial Intelligence are Challenging the Global Cybersecurity Workforce 2023. [Online]. https://www.isc2.org/research. (visited on 2025-01-06)

\bibitem{SP800}
NIST. NIST SP 800-53 Rev. 5, Security and Privacy Controlsfor Information Systems and Organizations. [Online]. https://csrc.nist.gov/pubs/sp/800/53/r5/upd1/final. (visited on 2024-08-20)

\bibitem{DataProvenance}
Pan, Bofeng, et al. "Data Provenance in Security and Privacy" ACM Comput. Surv. 2023

\bibitem{CSS}
Taishin Saito,Masaki Hashimoto,Kuniyasu Suzaki. "	Linking Benign Activities in Information Systems with DARPA Data" 	Computer Security Symposium 2024. 2024

\bibitem{PROGRAPHER}
Fan Yang, et al. "PROGRAPHER: An Anomaly Detection System based on Provenance Graph Embedding" 32nd USENIX Security Symposium (USENIX Security 23), 2023

\bibitem{SLEUTH}
Md Nahid Hossain, et al. "SLEUTH: Real-time Attack Scenario Reconstruction from COTS Audit Data" 26th USENIX Security Symposium (USENIX Security 17), 2017

\bibitem{UNICORN}
 Xueyuan Han, et al. "UNICORN: Runtime Provenance-Based Detector for Advanced Persistent Threats" 3 Network and Distributed Systems Security (NDSS) Symposium 2020, 2020

 \bibitem{ATLAS}
 Abdulellah Alsaheel, et al. "ATLAS: A Sequence-based Learning Approach for Attack Investigation" 30th USENIX Security Symposium (USENIX Security 21), 2021


\bibitem{Muhammad}
M. A. Inam, et al. "SoK: History is a Vast Early Warning System: Auditing the Provenance of System Intrusions" 2023 IEEE Symposium on Security and Privacy (SP), San Francisco, CA, USA, 2023

\bibitem{p-EDR}
Zipperle, Michael, et al. "Provenance-based Intrusion Detection Systems: A Survey" ACM Comput. Surv. 2023

\bibitem{p-EDR-interview}
Dong, Feng, et al. "Are we there yet? An Industrial Viewpoint on Provenance-based Endpoint Detection and Response Tools" Proceedings of the 2023 ACM SIGSAC Conference on Computer and Communications Security. 2023

\bibitem{Backtracker}
King, Samuel T. and Chen, Peter M. "Backtracking intrusions" SIGOPS Oper. Syst. Rev. 2003

\bibitem{LogGC}
Lee, Kyu Hyung, et al. "LogGC: garbage collecting audit log" Proceedings of the 2013 ACM SIGSAC Conference on Computer \& Communications Security. 2013

\bibitem{CPR}
Xu, Zhang, et al. "High Fidelity Data Reduction for Big Data Security Dependency Analyses" Proceedings of the 2016 ACM SIGSAC Conference on Computer and Communications Security. 2016

\bibitem{DEPIMPACT}
Pengcheng Fang, et al. "{Back-Propagating} System Dependency Impact for Attack Investigation" 31st USENIX Security Symposium (USENIX Security 22). 2022 

\bibitem{NODOZE}
Hassan, Wajih, et al. "NoDoze: Combatting Threat Alert Fatigue with Automated Provenance Triage" Proceedings 2019 Network and Distributed System Security Symposium(NDSS). 2019

\bibitem{nodlink}
Li, Shaofei, et al. "NODLINK: An Online System for Fine-Grained APT Attack Detection and Investigation" Proceedings 2024 Network and Distributed System Security Symposium(NDSS). 2024

\bibitem{nodlink-git}
"NODLINK". [Online]. https://github.com/PKU-ASAL/Simulated-Data. (visited on 2025-01-06)

\bibitem{suetugu}
Suetsugu Nobuki, et al. "Investigation of linking and visualization method for malicious activities towards intrusion detection" 2020 Information Processing Society of Japan. 2020

\bibitem{miyasaka}
Tsuyoshi Miyasaka, et al. "Linking and Visualizing Malicious Activities for Intrusion Detection" Computer Security Symposium 2021. 2021

\bibitem{Threatrace}
Wang, Su, et al. "THREATRACE: Detecting and Tracing Host-Based Threats in Node Level Through Provenance Graph Learning" IEEE Transactions on Information Forensics and Security. 2022

\bibitem{Threatrace-git}
"THREATRACE". [Online]. https://github.com/threaTrace-detector/threaTrace. (visited on 2025-01-06)

\bibitem{fasttext}
Piotr, Bojanowski, et al. "Enriching Word Vectors with Subword Information" Transactions of the Association for Computational Linguistics. 2017

\bibitem{Darpa-TC}
DARPA. TC: Transparent Computing [Online]. https://www.darpa.mil/research/programs/transparent-computing. (visited on 2025-01-06)

\bibitem{Darpa-E3}
"Transparent Computing Engagement 3 Data". [Online]. https://github.com/darpa-i2o/Transparent-Computing/blob/master/README-E3.md. (visited on 2025-01-06)

\bibitem{Darpa-E5}
"Transparent Computing Engagement 5 Data". [Online]. https://github.com/darpa-i2o/Transparent-Computing. (visited on 2025-01-06)

\bibitem{Darpa-E3GroundTruth}
Kudu Dynamics. TA5.1 Ground Truth Report Engagement 3 . [Online]. https://drive.google.com/drive/folders/1ATro9\_PaoNlg376yA\_moI1MbJGF-\_HaV. (visited on 2025-01-06)

\bibitem{Darpa-E5GroundTruth}
Kudu Dynamics. TA5.1 Final Report Engagement 5 . [Online]. https://drive.google.com/drive/folders/19rZOi3EDyv8Oa9yv8Tp7lRgL7F8b5fQ7. (visited on 2025-01-06)


\end{thebibliography}
